\RequirePackage[2020-02-02]{latexrelease}
\documentclass[lettersize,journal]{IEEEtran}
\usepackage{amsmath,amsfonts}
\usepackage{algorithm,algorithmic}
\usepackage{array}
\usepackage[caption=false,font=normalsize,labelfont=sf,textfont=sf]{subfig}
\usepackage{textcomp}
\usepackage{stfloats}
\usepackage{url}
\usepackage{verbatim}
\usepackage[switch, mathlines]{lineno}
\usepackage{graphicx}
\usepackage{graphics}
\def\BibTeX{{\rm B\kern-.05em{\sc i\kern-.025em b}\kern-.08em
    T\kern-.1667em\lower.7ex\hbox{E}\kern-.125emX}}
\usepackage{balance}
\usepackage{cite}
\usepackage{hyperref}
\usepackage{xcolor}
\usepackage{multirow}
\usepackage[printwatermark]{xwatermark}
\usepackage{tabularx}
\usepackage{transparent}
\usepackage{colortbl,booktabs}

\begin{document}
\title{Deconstructing Pedestrian Crossing Decision-making in Interactions with  Continuous Traffic: an Anthropomorphic Model}

\author{Kai Tian, Gustav Markkula, Chongfeng Wei \IEEEmembership{Member, IEEE}, Yee Mun Lee, Ruth Madigan, Toshiya Hirose, Natasha Merat and Richard Romano
\thanks{Manuscript received. Corresponding author: Kai Tian}
\thanks{Kai Tian, Gustav Markkula, Yee Mun Lee, Ruth Madigan, Natasha Merat and Richard Romano are with Institute for Transport Studies, University of Leeds, Leeds, United Kingdom, LS2 9JT (E-mail: tskt@leeds.ac.uk; G.Markkula@leeds.ac.uk; Y.M.Lee@leeds.ac.uk; R.Madigan@leeds.ac.uk; N.Merat@its.leeds.ac.uk; R.Romano@leeds.ac.uk).}
\thanks{Chongfeng Wei is with School of Mechanical and Aerospace Engineering, Queen's University Belfast, Belfast, United Kingdom, BT7 1NN (E-mail: C.Wei@qub.ac.uk).}
\thanks{Toshiya Hirose is with Department of Engineering Science and Mechanics, Shibaura Institute of Technology, Tokyo, Japan (E-mail: hiroset@shibaura-it.ac.jp). This work was supported by the UK Engineering and Physical Sciences Research Council under grant EP/S005056/1.}
}

\maketitle
\thispagestyle{empty}
\pagestyle{empty}

\begin{abstract}
As safe and comfortable interactions with pedestrians could contribute to automated vehicles' (AVs) social acceptance and scale, increasing attention has been drawn to computational pedestrian behavior models. However, very limited studies characterize pedestrian crossing behavior based on specific behavioral mechanisms, as those mechanisms underpinning pedestrian road behavior are not yet clear. Here, we reinterpret pedestrian crossing behavior based on a deconstructed crossing decision process at uncontrolled intersections with continuous traffic. Notably, we explain and model pedestrian crossing behavior as they wait for crossing opportunities, optimizing crossing decisions by comparing the visual collision risk of approaching vehicles around them. A collision risk-based crossing initiation model is proposed to characterize the time-dynamic nature of pedestrian crossing decisions. A simulation tool is established to reproduce pedestrian behavior by employing the proposed model and a social force model. Two datasets collected in a CAVE-based immersive pedestrian simulator are applied to calibrate and validate the model. The model predicts pedestrian crossing decisions across all traffic scenarios well. In particular, by considering the decision strategy that pedestrians compare the collision risk of surrounding traffic gaps, model performance is significantly improved. Moreover, the collision risk-based crossing initiation model accurately captures the timing of pedestrian crossing initiations within each gap. This work concisely demonstrates how pedestrians dynamically adapt their crossings in continuous traffic based on perceived collision risk, potentially providing insights into modeling coupled human-AV interactions or serving as a tool to realize human-like pedestrian road behavior in virtual AVs test platforms.

\end{abstract}


\section{Introduction}

\IEEEPARstart{C}{ontinued} advances in vehicle automation have brought us great anticipation that society will adopt highly automated vehicles (AVs) in the near future. However, this vision faces many unresolved challenges. One of them is to achieve smooth interaction between AVs and other road users. The consensus suggests that in the transition from manual to fully automated driving, there will be mixed traffic with AVs and other road users on the road\cite{palmeiroInteractionPedestriansAutomated2018}. A typical case is the expansion of the deployment of AVs from a few confined areas of low risk to other road users to a range of operational design domains, which could inevitably increase conflicts with other road users\cite{connectedcooperative}. Failures in interactions between AVs and other road users may hinder the large-scale adoption and social acceptance of AVs \cite{markkulaDefiningInteractionsConceptual2020,rasouliAutonomousVehiclesThat2019}. This, therefore, leads to the research context of this study, which is to promote safe and smooth communication and interaction in traffic \cite{palmeiroInteractionPedestriansAutomated2018,markkulaDefiningInteractionsConceptual2020,rasouliAutonomousVehiclesThat2019}. Pedestrians are generally regarded as the most vulnerable road users in modern transport systems, due to the lack of protective equipment and slow movement compared to other road users \cite{elhamdaniPedestrianSupportIntelligent2020}. Given that pedestrians' actions and intentions are nondeterministic, and the diversity and dynamism of their behavior, moving through this complicated environment is a challenge for AVs\cite{domeyerDriverpedestrianPerceptualModels}. Moreover, AVs' own behavior can also affect pedestrian road behavior, which introduces further uncertainties into interactions. In particular, the issues mentioned above become more pronounced at uncontrolled intersections where pedestrian behavior is more unpredictable, and safety problems are more common than on other controlled road sections, as there are no traffic signals to coordinate the interaction process \cite{zhaoGapAcceptanceProbability2019}. Additionally, most existing automated driving systems regard the driving task as a pure collision-free motion planning problem and view pedestrians in some contexts as rigid road obstacles, instead of social beings \cite{elhamdaniPedestrianSupportIntelligent2020,schneemannAnalyzingDriverpedestrianInteraction2016}. 

Against the above background, if AVs cannot properly understand the behavior of pedestrians, they may not improve traffic efficiency and safety as expected, but rather increase traffic dilemmas and additional issues\cite{millard-ballPedestriansAutonomousVehicles2018}. Accordingly, much attention has been drawn to one pressing issue, namely computational models for pedestrian road behavior, \cite{pekkanen2021variable,gilesZebraCrossingModelling2019,domeyerDriverpedestrianPerceptualModels,zhangPedestrianPathPrediction2020,predhumeau2021pedestrian}, which may help AVs to better anticipate pedestrian intentions or serve as a tool to implement realistic pedestrian behavior in simulated scenarios, and thus be used in the validation and development of AVs\cite{markkulaDefiningInteractionsConceptual2020,rasouli2022intend}. Existing computational models for pedestrian behavior, particularly for pedestrian road-crossing decisions have been developed based on a wide range of theories and hypotheses, such as the cognitive models \cite{markkulaModelsHumanDecisionmaking2018,pekkanen2021variable}, data-driven approaches \cite{volzInferringPedestrianMotions2018}, discrete choice models\cite{zhangPedestrianPathPrediction2020}, as well as game theoretical models \cite{camaraEmpiricalGameTheory2018}. However, those approaches have not yet bridged several gaps, as identified and discussed below.

Firstly, most of these approaches are rarely based on specific behavioral or psychological theories, such as pedestrian visual perception. Instead, external physical factors, like time to collision (TTC), have been often used. For example, \cite{zhangAnalysisPedestrianStreetCrossing2020,fuNovelFrameworkEvaluate2018} developed a pedestrian crossing decision-making model based on the vehicle deceleration distance. \cite{zhu2021novel,rasouli2022intend} applied a minimum TTC as the threshold for pedestrian crossing decisions. Although TTC or distance from the vehicle has become the most used decision cue in crossing decision models\cite{zhangAnalysisPedestrianStreetCrossing2020}, growing evidence has shown that the impacts of vehicle kinematics on pedestrians are multi-dimensional. For instance, at the same TTC condition, a higher vehicle speed induces more pedestrians to cross the street compared to a lower one\cite{lobjoisAgerelatedDifferencesStreetcrossing2007}. Therefore, the TTC or distance may not properly carry the risk information that pedestrians may perceive. As our previous research has shown, pedestrian crossing behavior is highly correlated with their perceived visual cues \cite{tian2022explaining}. Hence, existing models lack effort in characterising pedestrian perceived information, e.g., anthropomorphic visual cues\cite{pekkanen2021variable,palmeiroInteractionPedestriansAutomated2018}.

Moreover, few computational models specifically characterize pedestrian decisions in the traffic flow scenario. In real situations, pedestrians usually face a fleet of vehicles and accept one traffic gap after rejecting some gaps. Thus, the decision-making in continuous traffic may not only be based on the collision risk, but also involve many trade-offs between safety and time efficiency\cite{suchaPedestriandriverCommunicationDecision2017}. Several previous studies indicated that with the increased waiting time, pedestrians tended to accept crossing opportunities with higher risk\cite{zhaoGapAcceptanceProbability2019}. \cite{rasouli2022intend} developed a model which hypothesized that pedestrians would change their obedience to the law when they waited a long time. However, there is much evidence that pedestrians who tended to wait were more cautious and less likely to accept risky gaps\cite{lobjoisEffectsAgeTraffic2013,tian_impacts_2022,yannisPedestrianGapAcceptance2013}. A meta-study uncovered these conflicting results and noted that there was insufficient evidence to support a linear relationship between waiting times and pedestrians risking crossing the street\cite{theofilatos2021cross}. On the one hand, the available findings support that pedestrians may dynamically adjust their crossing decision-making strategies in continuous traffic. On the other hand, it is unreasonable to assume that pedestrians always tend to accept more dangerous crossing opportunities as waiting time increases. Instead, we should treat each case on its own merits.  Therefore, it is necessary to look into the details of pedestrian crossing behavior when interacting with traffic flow.

Finally, very limited models pay attention to the time dynamic of pedestrian crossing decision-making. According to the cognitive decision-making theory, pedestrian crossing initiation time (or onset time) is a variable due to the noisy evidence in the human cognitive system \cite{markkulaAccumulationContinuouslyTimevarying2021}. In addition, it has been shown that pedestrian crossing initiation time can be affected by many factors. For instance, pedestrians may initiate quickly when facing a vehicle with a low speed\cite{lobjoisAgerelatedDifferencesStreetcrossing2007} or with a small time gap from the approaching vehicle\cite{kalantarovPedestriansRoadCrossing2018}. Accordingly, existing empirical observations highlight the time-dynamic nature of pedestrian crossing decision-making. Recently, a class of emerging models\cite{markkulaModelsHumanDecisionmaking2018,gilesZebraCrossingModelling2019,pekkanen2021variable}, namely the evidence accumulation model, detailed model pedestrian crossing decisions and their timing by simulating the cognitive process underlying crossing decision-making. However, given the complexity of those models, they focused more on the details of the cognitive process, and it is unclear whether it would be feasible to extend them to cover additional factors, such as vehicle kinematics.

Regarding the above discussion, several research questions in existing computational models of pedestrian crossing behavior can be summarised: 
\begin{itemize}
    \item There is a lack of computational models that characterize  pedestrian crossing decisions based on anthropomorphic behavioral theory.
    \item The decision pattern of pedestrians crossing the road when interacting with the traffic flow remains unclear.
    \item There is a lack of computational models that concisely consider the time-dynamic nature of road crossing decisions and relate them to vehicle kinematics.
\end{itemize}

In this study, a decision-making model for pedestrians interacting with continuous traffic at uncontrolled intersections is proposed to solve the above questions. The main contributions of this paper are as follows: 
\begin{itemize}
   \item We formally apply our findings \cite{tian2022explaining} and extend it to a relatively complex traffic scenario, demonstrating that pedestrian crossing decisions are dynamic and intrinsically linked to their perceived collision risk. Specifically, a visual collision risk model is introduced as the main decision cue accounting for pedestrian crossing decisions. Moreover, a novel decision strategy is proposed to interpret pedestrian crossing decisions in continuous traffic flow. In addition, a crossing initiation time model is developed and associated with the collision cue model to account for the pedestrian dynamic crossing initiation time.
   \item Two different datasets collected in a highly immersive pedestrian simulator are applied to calibrate and validate the model.
   \item A simulation tool is established to reproduce pedestrian crossing decisions in a customized traffic scenario based on the proposed model.
\end{itemize}

\section{Methodology}
\begin{figure}[htpb]
      \centering
     \includegraphics[scale=1]{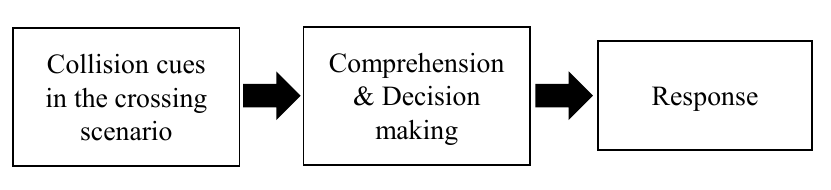}
      \caption{A simplified framework for pedestrians road-crossing decision-making process.}
      \label{figlabel1}
\end{figure}

\begin{figure*}[t]
      \centering
     \includegraphics[scale=0.78]{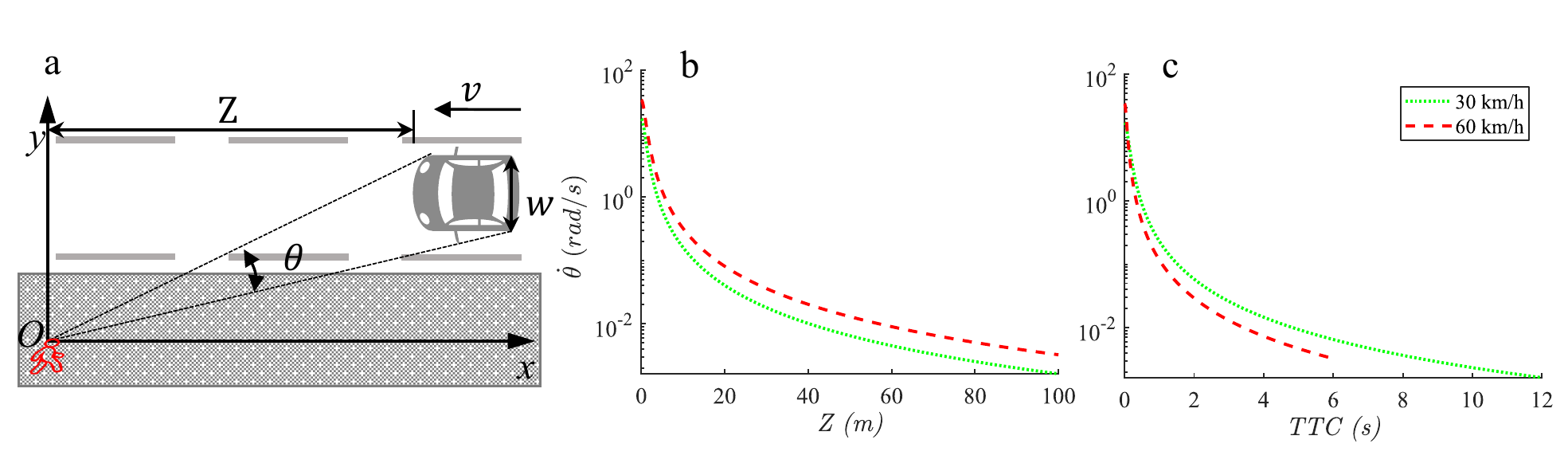}
      \caption{(a) Visual collision cue model in road crossing scenario. Collision cues are as a (b) function of distance from and speed of the vehicle or (c) TTC from and speed of the vehicle.}
      \label{figlabel2}
\end{figure*}

\subsection{Deconstructing the crossing decision-making process} 
During the decision-making process for road-crossing, several cognitive stages may be involved to establish pedestrian situation awareness\cite{palmeiroInteractionPedestriansAutomated2018,coeugnetRisktakingEmotionsSociocognitive2019}. Normally, pedestrian perceived collision cues are the basis of their decisions, which contain vehicle distance, speed, TTC, and more. Based on those visual cues, pedestrians comprehend traffic situations and decide whether to cross the road or not by combining some prior knowledge and strategies. Finally, there is a reaction process before pedestrians start to move. Therefore, according to the deconstructed three-stage cognitive process, we propose a collision cue-based framework for road-crossing decision-making tasks (Fig. \ref{figlabel1}), assuming that the crossing decision-making model consists of three constituent parts: visual collision cue, decision, and crossing initiation. 

\subsection{Visual collision cue model}

Modeling pedestrian-vehicle interaction is challenging, partly because existing pedestrian models lack psychological underpinnings. According to psychological theory, when moving through the environment, people rely on their visual perception of the space around them \cite{gibsonEcologicalApproachVisual2014,deluciaCriticalRolesDistance2008}. The road crossing task is a typical case that highly demands pedestrians to use visual cues to evaluate the collision risk from approaching vehicles and guide their movements. Relevant behavioral research has shown that the human visual system is sensitive to changes in some visual cues, which may be the source of collision perception. Specifically, one group of cues may provide reliable collision time information, such as Tau\cite{leeTheoryVisualControl1976}. Other cues, like visual angle and its first temporal derivative\cite{deluciaCriticalRolesDistance2008}, effectively translate motion information into visual cues through images that expand on the retina. Although most daily naturalistic road crossings involve all of the above visual cues (and possibly others), Delucia\cite{deluciaCriticalRolesDistance2008} has suggested that humans may rely on collision time-related cues when the scenarios include robust optical information or occur at a near distance. Conversely, when the optical information in the task is impoverished or occurs at a long distance, the visual angle and its first temporal derivative may play a dominant role. In light of this conceptual framework, we have previously identified that the first temporal derivative of visual angle, $\dot{\theta}$, is a critical collision cue for making crossing decisions at uncontrolled intersections. We have demonstrated that $\dot{\theta}$ not only well explains pedestrian crossing decisions across a wide range of traffic scenarios from two different datasets, but also reasonably characterizes the impacts of vehicle speed and traffic gap on pedestrians \cite{tian2022explaining}. Therefore, in this study, we formalized the pedestrian crossing decision model based on our previous findings. Typically, $\dot{\theta}$ refers to the change rate of the visual angle subtended by an approaching vehicle, $\theta$, (Fig. \ref{figlabel2}a)\cite{gibsonEcologicalApproachVisual2014}. The following equations specify its physical model:

\begin{equation}\label{1}
\theta=2 \tan ^{-1} \frac{w}{2 Z}  \Rightarrow  \dot{\theta}\left(Z, v, w\right)=\frac{w v}{(Z)^{2}+w^{2} / 4} 
\end{equation}

\noindent where $v$ denotes the vehicle speed, $Z$ and $w$ are the distance to and width of the vehicle. To better interpret the collision cue model, an example is shown in Fig. \ref{figlabel2}. Suppose that a vehicle ( $w=1.95$ m) approaches the pedestrian with two different constant speeds (30 km/h and 60 km/h) from 100 m. $\dot{\theta}$ is an approximately inversely exponential function of distance and TTC from the approaching vehicle (Fig. \ref{figlabel2}b, c), showing that $\dot{\theta}$ increases slowly at long distances and rapidly at close distances, which agrees qualitatively with the observation that pedestrians usually feel safe to cross for long distance or big time gap conditions but not when the vehicle is close \cite{lobjoisAgerelatedDifferencesStreetcrossing2007}. Further, it can be noticed that speed effects vary across distance (Fig. \ref{figlabel2}b) and TTC dimensions (Fig. \ref{figlabel2}c). When $\dot{\theta}$ is a function of distance and speed, it increases with speed, which is opposite to the results in Fig. \ref{figlabel2}c, suggesting that pedestrians may perceive a higher collision threat from the vehicle with higher speed at the same distance. However, the approaching vehicle with a slower speed gives pedestrians a bigger collision threat under the same TTC. The results tie well with the previous experimental observations on pedestrian crossing  behavior\cite{lobjoisAgerelatedDifferencesStreetcrossing2007,lobjoisEffectsAgingStreetcrossing2009,schmidtPedestriansKerbRecognising2009}. 

\begin{figure*}[t]
      \centering
     \includegraphics[scale=0.9]{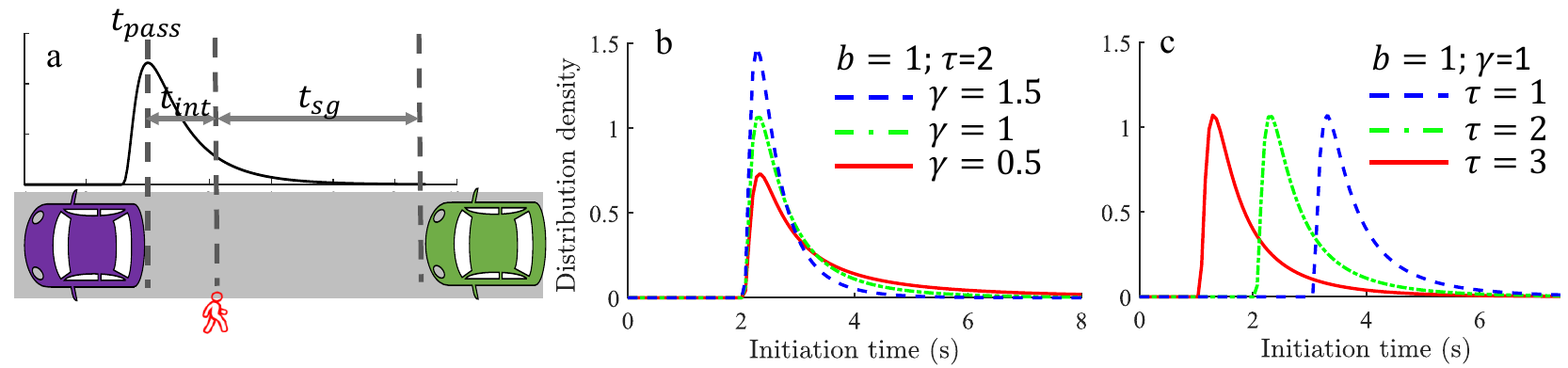}
      \caption{Illustration of the initiation model. (a) Initiation time $t_{int}$ is the duration between $t_{pass}$ and the time when the pedestrian start crossing. $t_{sg}$ denotes the actual gap to the approaching vehicle when pedestrians initiate. (b) The shapes of the initiation model by changing $\gamma$. (c) The positions of the initiation model by changing $\tau$.}
      \label{figlabel4}
\end{figure*}

\subsection{Decision model}
Regarding crossing decisions at uncontrolled intersections, pedestrians typically make crossing decisions by judging and selecting the appropriate gaps between two consecutive vehicles, called gap acceptance behavior\cite{zhaoGapAcceptanceProbability2019}. Our previous study has proven that $\dot{\theta}$ is significantly negatively correlated with pedestrian gap acceptance behavior, and a collision cue-based binary choice logit model predicts pedestrian gap acceptance well across different vehicle speeds and traffic gap experimental scenarios\cite{tian2022explaining}. Furthermore, evidence from experimental observations indicated that individuals' judgments toward traffic gaps are not necessarily entirely static over time, especially in traffic streams\cite{woodmanGapAcceptanceStudy2019,lobjoisEffectsAgeTraffic2013,tian_impacts_2022}. Due to certain learning or comparison strategies, pedestrians may estimate different utilities for the approaching vehicles with the same collision cues, thus adjusting their crossing decision to balance safety and efficiency. We, therefore, propose the following assumptions for the crossing decision-making in the traffic flow:

\noindent(i) Pedestrians make decisions mainly based on collision cues, i.e., $\dot{\theta}$, provided by approaching vehicles.
\\

\noindent(ii) Pedestrians are unwilling to accept the current gap with a collision cue equal to or greater than the maximum collision cue previously rejected. For example, if pedestrians reject a $0.02$ rad/s cue, they would be more likely to reject the same or bigger one upstream of traffic. The rule is given by:

\begin{equation}\label{2}
{X}_{1} =\left\{\begin{array}{l}
1,   \quad\dot{\theta}_{c} \geq \dot{\theta}_{mr} \\
0,   \quad\dot{\theta}_{c}<\dot{\theta}_{mr}
\end{array}\right. 
\end{equation}

\noindent where $X_{1}$ is the dummy variable for the rule.  $\dot{\theta}_{c} $ and $\dot{\theta}_{mr}$ represent collision cues for the current gap and maximum rejected gap, respectively.
\\

\noindent(iii) If pedestrians find that a gap next to the current gap has a smaller collision cue than the current gap, they may prefer to wait for this gap rather than accept a current gap with a greater collision threat, given the rule:

\begin{equation}\label{3}
{X}_{2} =\left\{\begin{array}{l}
1, \quad\dot{\theta}_{c} \geq \dot{\theta}_{f} \\
0, \quad\dot{\theta}_{c}<\dot{\theta}_{f}
\end{array}\right.
\end{equation}

\noindent where  $X_{2}$ is the dummy variable for the decision rule. $\dot{\theta}_{f}$ represents a collision cue of the gap following the current one.
\noindent Therefore, the utility function of the decision model is formulated as:

\begin{equation}\label{4}
V=\rho_{0} \ln (\dot{\theta})+\rho_{1} X_{1}+\rho_{2} X_{2}+\rho_{3} 
\end{equation}

 \noindent where $\rho_{0}$ to $\rho_{3}$ are estimated coefficients. In this study, every $\dot{\theta}$ only refers to the $\dot{\theta}$ value of the approaching vehicle at the time when the rear end of the previous vehicle just past the pedestrian (Fig. \ref{figlabel4}a). Regarding the $\ln$ transformation, we have previously proven that it can efficiently increase the accuracy of model fitting \cite{tian2022explaining}. Since crossing decisions at uncontrolled intersections are assumed to be a binary choice task, a logistic function is applied \cite{zhaoGapAcceptanceProbability2019}. Then, a decision model for crossing tasks in the traffic flow is given by:

\begin{equation}\label{5}
p(\dot{\theta},X_{1},X_{2})=\frac{1}{1+\exp \left(-V\right)} 
\end{equation}

\noindent where $p$ is the probability of the gap acceptance. The (\ref{5}) without the terms $X_{1}$ and $X_{2}$ degenerates to the model we proposed in \cite{tian2022explaining}.

\subsection{Crossing initiation model}\label{section2.3}

 In real traffic, the time at which pedestrians start to cross the road is a variable\cite{markkulaAccumulationContinuouslyTimevarying2021}. As illustrated in Fig. \ref{figlabel4}a, crossing initiation time, $t_{int}$, is typically defined as the duration between the time when the rear end of the previous car passes the pedestrians' position, $t_{pass}$, and the time when pedestrians start their movements\cite{lobjoisAgerelatedDifferencesStreetcrossing2007}. Emerging cognitive models\cite{markkulaAccumulationContinuouslyTimevarying2021,gilesZebraCrossingModelling2019,pekkanen2021variable} have shown that  the crossing initiation time distribution may arise from an underlying evidence accumulation process, but of a form that requires costly stochastic simulation of to estimate the distribution. However, the skewed, lognormal-like shape of the distribution is similar to those arising from simpler evidence accumulation processes, which can be written in a closed mathematical form, such as Ex-Gaussian, Shifted Wald (SW), and Weibull\cite{andersShiftedWaldDistribution2016}. Considering the similarities of those methods, we only apply the SW distribution instead of trying all of them. The SW distribution is a simple and concise distribution modeling tool, which can fully qualify the crossing initiation time distribution with three parameters: $b$ (deviation around the mode), $\gamma$ (tail magnitude) and $\tau$ (onset of the distribution). Its equation is defined as:

\begin{equation}\label{6}
\begin{gathered}
x \sim \operatorname{SW}(b, \gamma, \tau) \\
\Rightarrow \frac{b}{\sqrt{2 \pi(x-\tau)^{3}}} \cdot \exp \left(\frac{-[b-\gamma(x-\tau)]^{2}}{2(x-\tau)}\right)
\end{gathered}
\end{equation}

\noindent An illustration of the distributional effect that occurs by changing each of the $\gamma$ and $\tau$ parameters are shown in Fig. \ref{figlabel4} b and c. The tail becomes heavier as $\gamma$ decreases, (Fig. \ref{figlabel4}b). Changes in $\tau$ control the position of the distribution (Fig. \ref{figlabel4}c)\cite{andersShiftedWaldDistribution2016}. 

According to our assumptions in Fig. \ref{figlabel1}, the crossing initiation time model is affected by collision cues, so we define the initiation time model as follows:

\begin{equation}\label{7}
\begin{gathered}
t_{ int } \sim  \operatorname{SW}(b, \gamma, \tau) \\
\text { with } \gamma=\beta_{1} \ln (\dot{\theta})+\beta_{2} ; \tau=\beta_{3} \ln (\dot{\theta})+\beta_{4}
\end{gathered}
\end{equation}

\noindent where $t_{int }$ is the crossing initiation time. $\beta_{1}$ to $\beta_{4}$ are estimated coefficients. The idea behind these equations is that the strength of collision cues could affect the distribution pattern of pedestrian initiation time. For a more intensive collision threat, if pedestrians choose to cross, they tend to do so more quickly, so the distribution is concentrated and has a short tail. In contrast, when the collision threat is small, pedestrians tend to start crossing slowly, so the distribution is more likely to have a long tail\cite{lee2022learning}. Accordingly, the SW model is not only a practical distribution model but also provides notable psychological significance for our decision model. In addition, $b$ is assumed to be a coefficient not influenced by collision cues. Furthermore, since response time data are routinely assumed to be normally distributed in many studies \cite{lobjoisAgerelatedDifferencesStreetcrossing2007,oxleyCrossingRoadsSafely2005}, another crossing initiation time model based on the Gaussian distribution is proposed as a comparison to the SW model, defined as the following equations:

\begin{equation}\label{8}
\begin{gathered}
t_{int } \sim \mathcal{N}(\mu, \sigma), \\
\text { with } \mu=\beta_{1} \ln (\dot{\theta})+\beta_{2} ; \sigma=\beta_{3} \ln (\dot{\theta})+\beta_{4} 
\end{gathered} 
\end{equation}

\noindent where $\mu$ and $\theta$ are parameters of the Gaussian model, $\mathcal{N}$.

\subsection{Pedestrian road-crossing decision-making model in traffic flow}

Finally, a pedestrian road-crossing decision-making model based on the SW distribution in the traffic flow (SW-PRD) is then established by employing (\ref{5}) and (\ref{7}):
\begin{equation}\label{9}
\begin{aligned}
& f_{S W}(t_{\text {int }})=\sum_{n=1}^{N} P_{n} \cdot \operatorname{SW}\left( b, \gamma\left(\dot{\theta}_{n}\right), \tau\left(\dot{\theta}_{n}\right)\right) \\
& P_{n}=p\left(\dot{\theta}_{n}, X_{1, n}, X_{2, n}\right) \cdot\left(1-P_{n-1}\right) \\
& P_{0}=0
\end{aligned} 
\end{equation}

\noindent where $n$ is the position number of the gap in the traffic flow. ${\dot{\theta}}_{n}$, $X_{1,n}$ and $X_{2,n}$ represent the decision variables for the $n$th traffic gap. $P_{n}$ means the recursive probability that pedestrians accept the $n$th gap, which is calculated based on $p$ and $P_{n-1}$. Similarly, a road-crossing decision model based on Gaussian distribution (G-PRD) is given by:
\begin{equation}\label{10}
\begin{aligned}
& f_{G} (t_{\text {int }})=\sum_{n=1}^{N} P_{n} \cdot \mathcal{N}\left( \mu\left(\dot{\theta}_{n}\right), \sigma\left(\dot{\theta}_{n}\right)\right)
\end{aligned} 
\end{equation}

\subsection{Simulation tool}

In this subsection, an agent-based simulation tool is proposed using the established models to reproduce pedestrian crossing behavior at uncontrolled intersections with traffic flow. The framework mainly includes three parts: the decision model, environment model, and pedestrian kinematics model. Regarding the traffic environment, as the intersections on multi-lanes are often separated by refuges\cite{davies1999research}, pedestrians actually cross one lane at a time. Therefore, a single-lane road with an uncontrolled intersection is considered. On the other hand, the model is possibly extended to a multi-lane situation, but the impacts of refuges should be further considered \cite{zhang2017quantitative}. A fleet of vehicles travels on the lane at a constant speed, wherein the vehicle quantity, speed, and traffic gaps can be customized. Afterward, a basic social force model is applied as a pedestrian kinematics model\cite{farinaWalkingAheadHeaded2017}, which considers the driving force towards the destination and repulsive force from the boundary of the crosswalk. Finally, according to the information provided by the traffic environment and kinematics model, each pedestrian's road crossing decision is generated through PRD models. The detailed process of the simulation tool is provided in the supplementary file  (Appendix. \ref{FirstAppendix}). A demonstration video of the simulation tool is also provided. Please see the attachment.

\section{Model calibration and validation}
In this study, two empirical datasets collected in a simulated environment, i.e., a CAVE-based highly immersive pedestrian simulator, were applied to calibrate and validate the PRD models. The following sections provide detailed information on the two datasets, calibration, and validation methods.

\begin{figure}[htbp]
      \centering
     \includegraphics[scale=0.45]{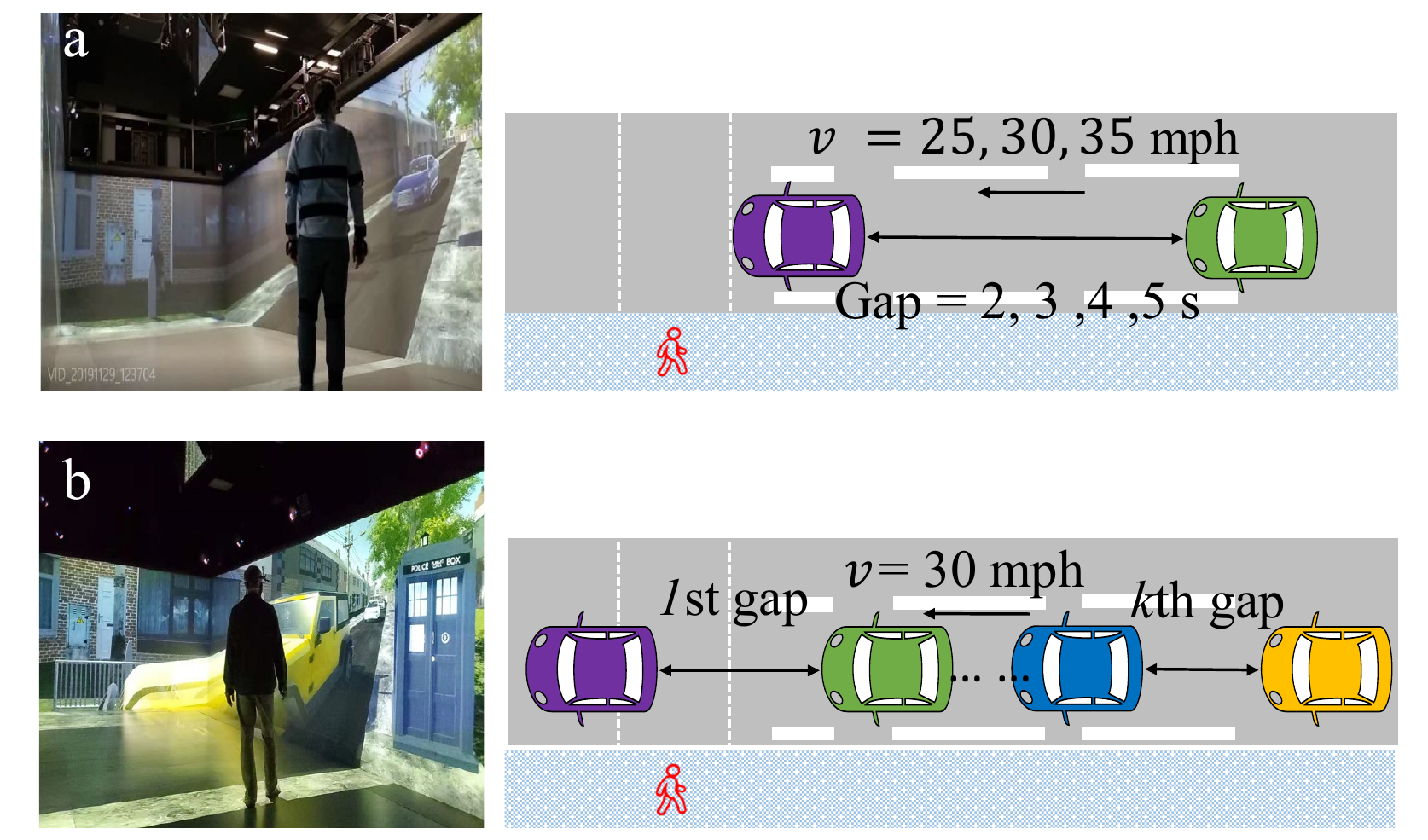}
      \caption{Schematic diagrams and photos of traffic scenarios in simulated experiments. The crossing scenarios and traffic of the (a) first dataset and (b) second dataset.}
      \label{figlabel6}
\end{figure}

\subsection{Empirical data}\label{3a}

\textit{Dataset one}. A virtual road scene with a 3.5 m wide single lane and 1.85 m wide pavement was created in the simulator. Two consecutive vehicles of 1.95 m in width were driven in the middle of the road at the same constant speed. Three vehicle speeds were selected, namely, 25 mph, 30 mph, or 35 mph. The first vehicle came into view 96 m away from the pedestrian, and the second vehicle maintained a specific time gap behind the first vehicle, i.e. 2 s, 3 s, 4 s, or 5 s (Fig. \ref{figlabel6}a). Sixty participants were instructed to cross the road between the two cars if they felt comfortable and safe to do so. Otherwise, they could reject the gap. Three experimental blocks were created, and each of the 12 scenarios (4 time gaps $\times$ 3 speeds) were presented in random order and repeated once in each experimental block. Therefore, each participant experienced 72 trials, and 4270 trials of data were obtained in total. 

The virtual environment and simulation process mentioned above were designed and controlled by the Unity3D platform. Internal code automatically recorded the positions and velocities of vehicles and participants on each time step. Two main metrics were applied: gap acceptance, $u$, and crossing initiation time, $t_{int}$. The gap acceptance data were the binary crossing decisions made by participants, i.e., $u=1$ means pedestrians accepted the gap, while 0 indicated rejected the gap. The crossing initiation time was defined as described in Section \ref{section2.3} and Fig.\ref{figlabel4}a.  For more detailed information about this dataset, please refer to \cite{lee2022learning}.

\textit{Dataset two}. To explore pedestrians' road crossing decisions in traffic flow, pedestrians were asked to cross a one-lane road with continuous traffic in the simulator (Fig.\ref{figlabel6}b). The size of time gaps between every two consecutive vehicles varied, which provided pedestrians with different opportunities to make crossing decisions (Fig.\ref{figlabel6}b). Four traffic scenarios with different sequences of gap sizes (in seconds) were designed as follows:
\begin{itemize}
\item Scenario one: 1 1 1 3 3 3 6 1 1 6;
\item Scenario two: 1 1 1 1 3 3 7 1 1 3 8;
\item Scenario three: 1 1 1 3 1 3 1 3 5 4 8;
\item Scenario four: 2 3 1 1 3 1 1 1 5 4 7;
\end{itemize}

Among these scenarios, the one-second and two-second time gaps between vehicles were considered dangerous crossing opportunities that very few pedestrians would accept. For the three-second and four-second gaps, decisions were expected to significantly differ between participants due to their heterogeneity (e.g., age and gender). The time gaps longer than four seconds were considered safe gaps that most pedestrians were expected to confidently accept. In all scenarios, a range of compact, midsize, van, and SUV vehicles were driven at 30 mph. Since the types of the approaching vehicle were randomly selected, in the analyses here, the width of the vehicle was calculated by averaging the width of all vehicles in the corresponding gap in each scenario. 60 participants completed four crossing tasks in any of the four scenarios and  repeated them once more (4 crossing tasks $\times$ 4 scenarios $\times$ 2 repetitions). We, therefore, collected data from 1920 trials. All the trials that participants experienced were in a randomized order. Similar to the first dataset, two main metrics were used: gap acceptance, $u$, and crossing initiation time, $t_{int}$. For more detailed information about this dataset, please refer to\cite{tian_impacts_2022}.

\subsection{Data processing and parameter estimation}
With regard to data processing, both datasets were divided into a training set and a validation set. Regarding dataset one, as controlled experimental variables were vehicle speed and time gap size, we separated the training and validation sets by choosing the data from different combinations of experimental variables (As illustrated in Section \ref{3a}, there were 12 different combinations). To have enough data in the training and validation sets, data from 10 combinations were grouped into the training set, while the rest of the data belonged validation set. Moreover, in order to make sure the validation data were sufficiently different, the 2 combinations are not adjacent to each other in terms of speed or time gap size. Accordingly, the validation set included data in 4 s 25 mph and 5 s 35 mph conditions, approximately accounting for $23\%$ of the initiation time data and $14\%$ of the gap acceptance data (The data size of the two metrics was not the same as there was no initiation time data for participants who rejected the gap). The remaining data of all other conditions were grouped into the training set. Similarly, with respect to dataset two, the data from traffic scenario four were used as the validation set, accounting for $24\%$ of gap acceptance data and $25\%$ of initiation time data. 

 A Maximum Likelihood Estimation (MLE) method was used to calibrate the parameters in the models. Firstly, regarding the decision model (\ref{5}), since it assumes that crossing decisions are drawn from a Bernoulli distribution, its likelihood function is given by:
\begin{equation}\label{11}
\begin{gathered}
\mathcal{L}_{1}(\omega)=\prod_{i=1}^{n} p\left(\Theta \mid \omega\right)^{u_{i}}\left(1-p\left(\Theta \mid \omega\right)^{1-u_{i}}\right) \\ 
\rho_{1}, \rho_{2}, \rho_{3}, \rho_{4} \in \omega \\
\dot{\theta}_{i}, X_{1,i}, X_{2,i} \in \Theta
\end{gathered}
\end{equation}
\noindent where $\omega$ includes all the estimated parameters $\rho_{1}, \rho_{2}, \rho_{3}, \rho_{4}$. $\Theta$ denotes $\dot{\theta}_{i}, X_{1,i}, X_{2,i}$ for the $i$th trial. $n$ is the size of the dataset. With respect to the initiation models, their likelihood functions are given by the following equations based on (\ref{7}) and (\ref{8}):
\begin{equation}\label{12}
\begin{gathered}
\mathcal{L}_{2}(\Delta)=\prod_{j=1}^{m} \operatorname{SW}\left(t_{int,j}, \dot{\theta_{j}}\mid \Delta\right)\\
\beta_{1}, \beta_{2}, \beta_{3}, \beta_{4}, b \in \Delta
\end{gathered}
\end{equation}

\begin{equation}\label{13}
\begin{gathered}
\mathcal{L}_{3}(\Delta)=\prod_{j=1}^{m} \mathcal{N}\left(t_{int,j}, \dot{\theta_{j}}\mid \Delta\right)
\end{gathered}
\end{equation}
\noindent where $\Delta$ is the summary of the estimated parameters of crossing initiation models. $t_{int,j}$ is the $j$th crossing initiation time data. The data size is $m$. According to the MLE method, the maximization problem is equivalent to minimizing the negative log-likelihood. Thus, the optimal estimations for parameters are achieved when negative log-likelihood functions are minimised, e.g., $-\ln \left(\mathcal{L}_{1}(\omega)\right)$. We applied a built-in 'fminuc' function in MATLAB to find the solution to the above minimization problems \cite{matlabVersion10R2021a2021}. 

Furthermore, there were some differences in the model estimates based on the two datasets. Firstly, since the traffic flow scenarios were not considered in dataset one, the models based on this dataset did not include the parameters $\rho_{1},\rho_{2}$. Regarding dataset two, for comparison purposes, we manipulated the SW-PRD model so that it had the proposed decision rules for traffic flow, whereas the G-PRD model did not. The estimated parameters based on the two datasets are presented in Table. \ref{table1} and Table. \ref{table2}. In addition, the parameters of the social force model are adopted from \cite{farinaWalkingAheadHeaded2017}.

\subsection{Validation methods}
After calibration, the predictions were compared with the validation set to verify the ability of the models. Two evaluation methods were applied to compare the performance of the proposed models, namely BIC and K-S test. The BIC is given by:
\begin{equation}\label{14}
\text{BIC}=k \ln (n)-2 \ln (L)
\end{equation}
\noindent where $k$ is the number of parameters in the model. $n$ is the size of the dataset. $L$ is the maximum likelihood. The preferred model is the one with the minimum BIC \cite{schwarzEstimatingDimensionModel1978}. K-S test is a nonparametric test, which is used to evaluate the goodness-of-fit of the predicted results by quantifying the distance between empirical and predicted distributions \cite{stephensEDFStatisticsGoodness1974}. The main equation of K-S test is:
\begin{equation}\label{15}
D_{n, m}=\sup \left|\boldsymbol{F}_{n}(x)-\boldsymbol{F}_{m}(x)\right|
\end{equation}
\noindent where $\sup$ denotes the supremum function. $\boldsymbol{F}_{n}(x)$ and $\boldsymbol{F}_{m}(x)$ are the distribution functions of the observed data and predicted result. $n$ and $m$ represent the size of the samples. The K-S test rejects the null hypothesis, i.e., two samples are drawn from the same probability distribution if $D_{n, m}$ is bigger than the selected threshold. In addition, the R-squared, $R^{2}$, and Root Mean Square Error (RMSE) are also used in the model discussion.

\begin{table}
\centering
\arrayrulecolor{black}
\caption{Calibration results of models based on dataset one}
\begin{center}
\tabcolsep 0.04in
\begin{tabular}{ccccc} 
\hline
\multirow{2}{*}{~ Parameter}    & \multicolumn{2}{l}{SW-PRD (Without flow)} & \multicolumn{2}{l}{G-PRD (Without flow)}  \\ 
\cline{2-5}
                                & Estimate & 95 \% C.I.                & Estimate & 95 \% C.I.                   \\ 
\hline
$\beta_{1}$ & 0.03      & {[}-0.19, 0.24]            & -0.03*     & {[}-0.05, -0.01]             \\ 

$\beta_{2}$ & 4.48*      & {[}3.35, 5.62]            & 0.15*      & {[} 0.07, 0.24]               \\ 

$\beta_{3}$ & -0.20*      & {[}-0.26, -1.78]          & -0.21*
& {[}-0.24, -0.18]             \\ 

$\beta_{4}$ & -2.11*    & {[}-2.43, 1.22]          & -0.76*     & {[}-0.91, -0.62]             \\ 

\textit{b}                      & 6.06*     & {[}4.43, 7.68]             & -         & -                            \\ 

$\rho_{0}$  & -2.14*     & {[}-2.28, -1.98]          & -2.14*     & {[}-2.28, -1.98]             \\ 
$\rho_{3}$  & -9.95*    & {[}-10.64, -9.26]        & -9.95*    & {[}-10.64, -9.26]           \\
LL                & \multicolumn{2}{l}{-108.43}          & \multicolumn{2}{l}{-176.69}             \\
BIC                              & \multicolumn{2}{l}{252.37}           & \multicolumn{2}{l}{381.79}              \\ 
\arrayrulecolor{black}\hline
\multicolumn{5}{l}{Note. LL: log-likelihood of the entire model, C.I.: confidence }\\
\multicolumn{5}{l}{interval, *: significant at a 5\% significance level}\\
\multicolumn{5}{l}{With/Without flow: consider/not consider decision strategies for traffic flow}                                                  
\end{tabular}
\end{center}
\label{table1}
\end{table}

\begin{table}
\centering
\arrayrulecolor{black}
\caption{Calibration results of models based on dataset two}
\begin{center}
\tabcolsep 0.04in
\begin{tabular}{ccccc} 
\hline
\multirow{2}{*}{~ Parameter}    & \multicolumn{2}{l}{SW-PRD (With flow)} & \multicolumn{2}{l}{G-PRD (Without flow)}  \\ 
\cline{2-5}
                                & Estimate & 95 \% C.I.                & Estimate & 95 \% C.I.                   \\ 
\hline
$\beta_{1}$ & 0.47*      & {[}0.29, 0.66]            & -0.05*     & {[}-0.06, -0.04]             \\ 

$\beta_{2}$ & 7.36*      & {[}6.15, 8.57]            & 0.01      & {[}-0.05, 0.07]               \\ 

$\beta_{3}$ & 0.04      & {[}-0.02, 0.10]          & -0.10*     & {[}-0.13, -0.09]             \\ 

$\beta_{4}$ & -1.41*     & {[}-1.70, -1.13]          & -0.59*     & {[}-0.68, -0.50]             \\ 

\textit{b}                      & 7.76*      & {[}5.6, 9.90]             & -         & -                            \\ 

$\rho_{0}$  & -2.92*     & {[}-3.16, -2.68]          & -3.31*     & {[}-3.55, -3.07]             \\ 

$\rho_{1}$  & -1.29*     & {[}-1.56, -1.02]          & -         & -                            \\ 

$\rho_{2}$  & -0.50*     & {[}-0.84, -0.15]          & -         & -                            \\ 

$\rho_{3}$  & -13.23*    & {[}-14.30, -12.16]        & -15.50*    & {[}-16.56, -14.46]           \\
LL(Decision model)               & \multicolumn{2}{l}{-1536.40}          & \multicolumn{2}{l}{-1672.50 }             \\
LL(CIT model)                & \multicolumn{2}{l}{-36.35}          & \multicolumn{2}{l}{-104.03}             \\
BIC                              & \multicolumn{2}{l}{3218.40}           & \multicolumn{2}{l}{3600.40}              \\ 
\arrayrulecolor{black}\hline
\multicolumn{5}{l}{Note. LL(Decision model/CIT model): log-likelihoods of decision models}  \\
\multicolumn{5}{l}{/crossing initiation time models}   
\end{tabular}
\end{center}
\label{table2}
\end{table}

\begin{figure*}[t]
      \centering
     \includegraphics[scale=1.1]{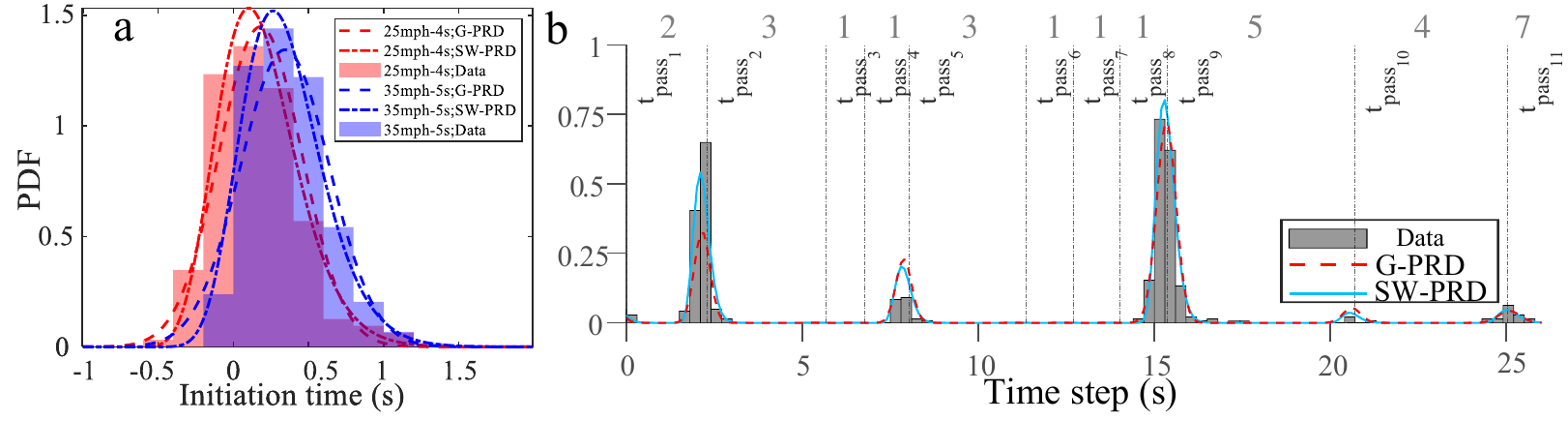}
      \caption{ Validation results. Probability density functions and data based on datasets (a) one and (b) two. The vertical dash-dotted lines in (b) indicate the time when the rear end of the vehicle passes the pedestrian's position. The size of the time gap (in seconds) between every two vehicles is indicated at the top of the diagram.}
      \label{figlabel7}
\end{figure*}

\section{Results and Analysis}
In this Section, we first discuss the calibration results of the SW-PRD and G-PRD models. Afterward, the validation results of the two models were compared using the BIC and K-S test. Finally, the model with better performance is compared to two entire datasets, and the reproduced crossing behavior patterns are discussed in detail. Additionally, regarding the first dataset, as it does not include the traffic flow scenario, we focus on the impacts of speed and time gap on pedestrian crossing behavior, while the effect of traffic is discussed using the results based on the second dataset.

\begin{table}[b]
\centering
\caption{Validation results of models based on dataset one}
\arrayrulecolor{black}
\begin{center}
\tabcolsep 0.04in
\begin{tabular}{cccccc} 
\hline
Condition & Model & LL & BIC   & K-S test score & P value  \\ 
\hline
25 mph 4 s    & SW-PRD & -23.08         & 71.47 & 0.06           & 0.56     \\
~             & G-PRD  & -27.28         & 74.82 & 0.10           & 0.08     \\
35 mph 5 s    & SW-PRD & -13.19         & 54.81 & 0.05           & 0.31     \\
~             & G-PRD  & -24.83         & 72.41 & 0.09           & 0.02*    \\ 
\hline                      
\end{tabular}
\end{center}
\label{table3}
\end{table}

\subsection{Calibration results}
 \textit{Dataset one}. The parameters of the SW-PRD and G-PRD models were calibrated using the first dataset. One thing to note is that as the first dataset did not include traffic flow scenarios, these two models thus did not implement decision strategies in traffic, which means $\rho_{1}$ and $\rho_{2}$ were not included in the models, and two decision models in the SW-PRD and G-PRD models were the same. The calibration results are shown in Table. \ref{table1}, where the maximum log-likelihood and BIC of the SW-PRD model based on the training set are -108.43 and 252.37, which are significantly better than those of the G-PRD model, i.e., -176.69 and 381.79, indicating that the SW-PRD model can better describe pedestrian crossing initiation time than the G-PRD model on the calibration set. Moreover, it can be found that the effect of $\rho_{0}$ is significantly negatively correlated with $\dot{\theta}$ ($\text{Est}.=-2.14, \text{C.I.}=[-2.28,-1.98]$), showing that pedestrian crossing gap acceptance decreases as the risk of collision increases. Additionally, the estimated effect of $\beta_{3}$ in the SW-PRD model is significantly correlated with $\dot{\theta}$ (Table. \ref{table1}), suggesting that pedestrian crossing initiation time is negatively related to the collision risk.

\textit{Dataset two}. The calibration results based on the second dataset are shown in  Table. \ref{table2}. As the SW-PRD model implemented the decision strategies in traffic flow, it included $\rho_{1}$ and $\rho_{2}$. However, the G-PRD model did not. Meanwhile, as both the decision model and initiation time model in the SW-PRD model and the SW-PRD model were different, we calculated the respective log-likelihood of the decision and initiation time models to facilitate the comparison of the results. Again, the SW-PRD model fits data better than the G-PRD model, where the SW-PRD model has larger log likelihoods for both the decision and crossing initiation time models, and its BIC is smaller than that of the G-PRD model. In particular, concerning the SW-PRD model, except for the significant effect of $\rho_{0}$ ($\text{Est}.=-2.92, \text{C.I.}=[-3.16,-2.68]$), $\rho_{1}$ and $\rho_{2}$ also significantly affect the pedestrian gap acceptance ($\text{Est}.=-1.29,\text{C.I.}=[-1.56, -1.02]; \text{Est.}=-0.50, \text{C.I.}=[-0.84, -0.15] $), consistent with our assumed crossing decision strategies in traffic flow. In addition, although the effect of $\beta_{3}$ in the SW-PRD model is not significant, the positive effect of $\beta_{1}$ reduces the tail magnitude of the distribution of crossing initiation time as $\dot{\theta}$ increases and thus can reduce pedestrians crossing initiation time.

\subsection{Validation results}
The calibration results indicate that the SW-PRD model fits the training sets better than the G-PRD model. In this section, the validation sets of two datasets are compared with the predicted results of two models.

\textit{Dataset one}. Regarding the validation results, as shown in Table. \ref{table3}, the SW-PRD model has better BIC values and K-S scores for all conditions. Specifically, in the 35 mph 5 s condition, the K-S test rejects the null hypothesis and indicates that the results of the G-PRD model are different from the observed data at a $5\%$ significance level. As shown in Fig. \ref{figlabel7}a, it can be found that the G-PRD model tends to overestimate the initial parts of the data, but the SW-PRD model does not.

\textit{Dataset two}. The predicted results are compared to the validation set of the second dataset. The log-likelihood of crossing initiation time models of SW-PRD  and G-PRD are presented separately for reasons explained previously (Table. \ref{table4}). Both SW-PRD and G-PRD models accurately capture the timing of pedestrian crossing decisions in the traffic flow, i.e., the peak location of the initiation time distribution ( Fig. \ref{figlabel7}b). The predicted peak shapes of both models are close to the data. However, the SW-PRD model has a relatively better performance than the G-PRD model because the log-likelihood of the crossing initiation time model for SW-PRD is bigger than the value for G-PRD in Table. \ref{table4}. The overall predictions of the SW-PRD model are closer to the data than these of the G-PRD model. Specifically,  the SW-PRD model has a better BIC value and log-likelihood than the G-PRD model (Table. \ref{table4}). Also, the K-S test supports that the predicted density function of the SW-PRD model is similar to the empirical distribution. In contrast, the predicted result of the G-PRD model is rejected by the K-S test at a $5\%$ significance level (Table. \ref{table4}). As shown in Fig. \ref{figlabel7}b, it can be found that consistent with the empirical data, the SW-PRD model predicts a decrease in the gap acceptance from the first 3 s gap (at $t_{pass_{2}}$) to the second 3 s gap (at $t_{pass_{5}}$). By contrast, the G-PRD model calculates a constant value for both 3 s gaps, resulting in a significant underestimation of gap acceptance in the first 3 s gap. In general, the SW-PRD model has better performance than the G-PRD model on the validation set of dataset two.

\begin{table}[htpb]
\centering
\caption{Validation results of models based on dataset two}
\arrayrulecolor{black}
\begin{center}
\tabcolsep 0.03 in
\begin{tabular}{cccccc} 
\hline
Model                 & LL & LL(CIT model) & BIC     & K-S
  test score & $p$ value  \\ 
\hline
SW-PRD                 & -578.37 &  -11.23     & 1193.10 & 0.08             & 0.10         \\ 
G-PRD                  & -707.53 &  -52.76    & 1444.10 & 0.16             & 0.001*        \\
\arrayrulecolor{black}\hline
\end{tabular}
\end{center}
\label{table4}
\end{table}

In the following sections, we discuss the predicted pedestrian crossing behavior patterns in detail by comparing predicted results with the two full datasets to provide a complete understanding of the proposed crossing decision-making model. Since SW-PRD performs better on all datasets than G-PRD, the SW-PRD model generates our results in the following sections.

\subsection{Dataset one: Speed and time gap effects}

\begin{figure*}[t]
\centering
     \includegraphics[scale=1.2]{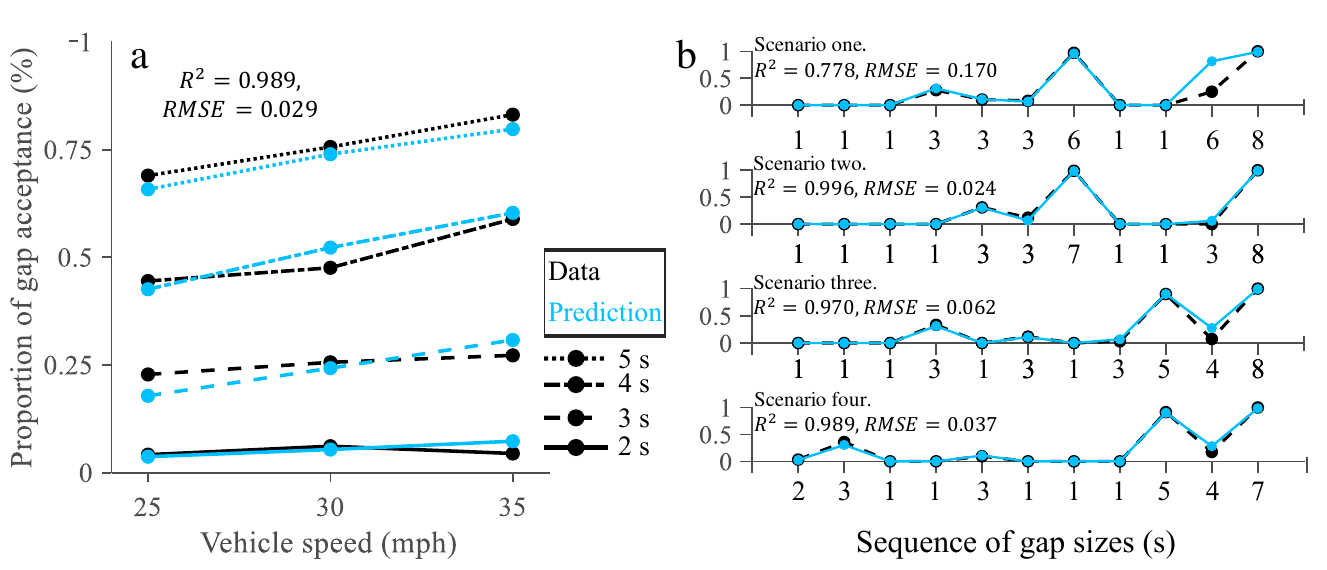}
      \caption{Predicted gap acceptance of the SW-PRD model for both datasets. The data and the predicted results are represented in black and blue respectively. (a) For dataset one, the proportion of gap acceptance is plotted as a function of vehicle speed and gap size (Gap sizes are indicated by different line styles). (b) For dataset two, the proportion of gap acceptance for each gap of each traffic scenario is presented.}
      \label{figlabel8}
\end{figure*}

 The SW-PRD model predictions of crossing gap acceptance for each speed and time gap condition are compared with the observed data in Fig. \ref{figlabel8}a. According to the empirical data, crossing gap acceptance increased with vehicle speed and traffic gap, aligning well with previous studies \cite{lobjoisAgerelatedDifferencesStreetcrossing2007,schmidtPedestriansKerbRecognising2009}. The SW-PRD model reproduces these behavioral patterns very well ($R^2=0.890$, $RMSE=0.050$), suggesting that pedestrians might adapt their crossing decisions based on the changes in collision cues.

\begin{figure}[htbp]
      \centering
     \includegraphics[scale=1]{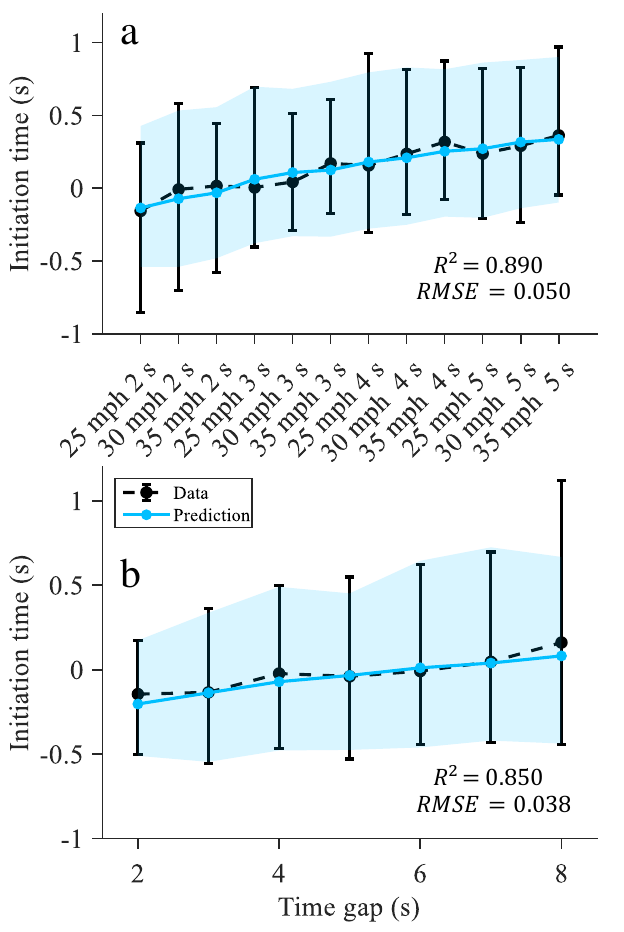}
      \caption{Predicted crossing initiation time of the SW-PRD model for both datasets. Error bars and the edge of blue areas indicate the  $2.5\%$ and $97.5\%$ percentiles of the data and predicted results. (a) For dataset one, the crossing initiation time is plotted as a function of vehicle speed and gap size. (b) For dataset two, the crossing initiation time is a function of gap size.}
      \label{figlabel9}
\end{figure}

Fig. \ref{figlabel9}a shows a comparison between the predicted crossing initiation time and observed data. In line with the literature, \cite{lobjoisEffectsAgingStreetcrossing2009}, the empirical data showed that pedestrian crossing initiation time correlated with vehicle kinematics, i.e., it decreased as traffic gaps and vehicle speeds decreased. This behavioral pattern can be understood as a distance-dependent phenomenon whereby a reduction in vehicle speed and time gap leads to a reduction in spatial distance, resulting in an increase in the perceived risk of collision \cite{tian2022explaining}. Hence, if pedestrians choose to cross, they tend to do so more quickly. Based on our modeling results, the proposed SW-PRD model captures this pattern with a good fit ($R^2=0.890$, $RMSE=0.050$), again indicating that visual collision cues are associated with pedestrian crossing behavior.

Moreover, a more detailed comparison between predictions and data is shown in Fig. \ref{figlabela2} in Appendix \ref{SecondAppendix}. It can be noticed that the SW-PRD model predicts pedestrian crossing behavior qualitatively and quantitatively. It not only describes the distributions of pedestrian crossing initiation along the time axis but also captures the variation in the mean crossing initiation time.

\subsection{Dataset two: Impacts of traffic flow}
Predicted gap acceptances of the SW-PRD model in the traffic flow are compared to the observed data in Fig. \ref{figlabel8}b. Firstly, it can be noticed that pedestrians in the traffic flow did not accept gaps of the same size equally. For instance, regarding the $4$th gap and the $5$th gap in traffic scenario one (The size of both traffic gaps is 3 s), the probability of crossing gap acceptance dropped significantly from $27.9 \%$ to $10.5 \%$. When pedestrians faced the $6$th gap, the decreasing trend became even stronger. The probability of crossing gap acceptance was $8.1 \%$, more than three times smaller than the value of the $4$th gap. Further looking at the predictions, the SW-PRD model reproduces this behavioral pattern across all traffic scenarios with reasonable goodness-of-fit Fig \ref{figlabel8}b).

Fig.\ref{figlabel9}b plots the predicted crossing initiation time as a function of the time gap and compares it with the observed data. The SW-PRD model fits the crossing  initiation time data well ($R^{2}=0.850$, $RMSE=0.038$). Consistent with empirical observations and similar to the first dataset \cite{kalantarovPedestriansRoadCrossing2018}, the SW-PRD model predicts a smaller initiation time as the time gap decreases, again suggesting that pedestrians attempted to compensate for crossing risk in unsafe traffic gaps by initiating faster.

Furthermore, as shown in Fig. \ref{figlabela3} in Appendix \ref{SecondAppendix}, detailed model predictions are compared with the observed data. Across all traffic scenarios, the SW-PRD model accurately predicts the level, shape and location of peaks of the crossing initiation time distribution, showing that the model has a good ability to characterize pedestrian crossing decisions in a continuous flow of traffic. 

\section{Discussion and conclusion}

This study demonstrates a novel approach to characterize pedestrian crossing decision-making at uncontrolled intersections with continuous traffic. We hypothesized that the crossing behavior could be understood as depending on three stages of information processing (perceive, decide, execute), and thus proposed a model with three corresponding constituent parts: visual collision cue, crossing decision, and crossing initiation. Following is a summary of the detailed research results.

In our previous study\cite{tian2022explaining}, we showed that the visual collision cue, $\dot{\theta}$, could capture the effects of vehicle kinematics on pedestrian crossing decisions in single gaps and explain why pedestrians tended to rely on distance from vehicles to make crossing decisions \cite{lobjoisAgerelatedDifferencesStreetcrossing2007,schmidtPedestriansKerbRecognising2009}. In this study, this finding is formally applied to model crossing decisions and extended to a more complicated traffic scenario, i.e., a continuous flow of traffic. The modeling results support that $\dot{\theta}$ is capable of characterizing the risk perceived by pedestrians, at least at uncontrolled intersections with constant speed traffic. 

Moreover, regarding our third hypothesis, i.e., pedestrian crossing initiation is time-dynamic and influenced by vehicle kinematics, we relate the proposed crossing initiation time model to $\dot{\theta}$. The modeling results support our hypothesis and show that pedestrians dynamically adjust their initiation time based on vehicle kinematics. Both the SW and Gaussian distributions can reasonably describe pedestrian initiation time, whilst the SW distribution has relatively better goodness-of-fit than the Gaussian distribution, which further indicates that the distribution of crossing initiation time is right-skewed. 

Notably, to accurately reproduce pedestrian crossing behavior in continuous traffic flow,  we further hypothesize that pedestrians compare the risks of the gaps around them before making decisions, which is supported by the fact that the proposed crossing decision strategy for continuous traffic scenarios significantly improves the performance of the model. The study thus concludes with the following findings. Firstly, pedestrians may have a reduced tendency to accept a gap if they see an upcoming larger gap. Secondly, pedestrians may have a greater tendency to reject a gap if they have already rejected a gap of that size or larger. Although no other studies have yet found these patterns of crossing behavior, some empirical observations provide indirect support. \cite{kittelson1991delay} showed that drivers who rejected the bigger traffic gap tended to incur a longer delay. \cite{yannisPedestrianGapAcceptance2013} indicated that pedestrians who tended to reject the crossing opportunities would be more cautious and tend to accept longer gaps. Moreover, \cite{lobjoisEffectsAgeTraffic2013} found that pedestrians who missed the first opportunity to cross the road would not compensate for their loss by accepting a shorter second opportunity to cross the road. The above studies reinforce our hypothesis that pedestrians who tend to wait for safer crossing opportunities are more cautious and more likely to optimize their crossing strategies by comparing crossing opportunities. Unlike several previous studies, which simply assumed pedestrians tend to accept smaller gaps with the increase in waiting time \cite{zhaoGapAcceptanceProbability2019,rasouli2022intend}, the novelty is that we show that there may be other patterns in pedestrian crossing behavior in terms of waiting for the crossing opportunity, which may provide an explanation for the non-significant effect of waiting time on pedestrian crossing decisions found in the meta-study \cite{theofilatos2021cross}. Furthermore, this finding is interesting in that it reminds us that there may be a complex changing pattern in pedestrians' strategy toward waiting for crossing opportunities. Future research can further attempt to disentangle the effects of waiting time and traffic flow.

Overall, this work provides a new concept that pedestrian crossing decisions are dynamic and intrinsically closely linked to their perceived collision risk, and can be reinterpreted through a three-stage crossing decision-making process. The proposed model shows good predictive performance in different simulator datasets, and it could therefore be interesting to test the model on naturalistic traffic datasets as a next step. Furthermore, the idea of the deconstructed process may drive further study to involve more complicated perceptual, decision, and initiation models. 

Regarding the practical implications of this study, there are many possible ways to extend these concepts and models to further improve research in pedestrian-AV interactions. First, as an increasing number of studies have been keen on using pedestrian behavior models to promote safe and efficient interactions\cite{camaraPedestrianModelsAutonomous}, the proposed decision model may provide predictive information to help automated driving systems to better anticipate pedestrian crossing intentions and initiations. Early work is emerging where researchers are attempting to plan and coordinate the actions of AVs and pedestrians toward common goals by considering the visual collision risk of pedestrians\cite{domeyerDriverpedestrianPerceptualModels}. 
Another possible application case is future traffic scenarios involving AV platoons and pedestrians, where AV platoons may need to take into account the dynamic pedestrian crossing decisions along the length of the platoon and adopt the decision strategy of each AV. Moreover, there is an urgent need to train and evaluate AVs to perform well also in safety-critical interactions with human road users. However, due to the low frequency of critical traffic scenarios in real life, i.e., the corner case, and safety reasons, both academia, and industry have agreed on using simulation methods as a complementary way to validate AVs. Reliable simulation results rely on the behavioral authenticity of simulated road users \cite{rasouli2022intend}. Hence, another practical significance of this study is that the model can serve as a module in the microscopic transport simulation tools or virtual testing platforms to realize naturalistic pedestrian road crossing decisions. 

However, several limitations of this study need to be addressed in the future. Since the results and model cover only scenarios with single-lane, constant-speed traffic, the model cannot be directly generalized to other scenarios without further development. For example, in situations with yielding vehicles, the collision cue model used in this study alone may not provide sufficient information to model crossing decisions. In addition, compared to the crossing behavior in pedestrian simulators, in real traffic, pedestrians can flexibly adjust their behaviors and be affected by many potential factors. The pedestrian simulator allows exact experimental control of conditions but, therefore, naturally provides a less variable environment, and the virtual nature of the task may also affect the observed behavior. Hence, an important future work should apply the model to a reliable naturalistic dataset. Furthermore, the model is developed based on current theories of human collision and does not assert that pedestrians exactly use the applied visual cues and perception strategy. As collision perception theory is further developed, the model can be improved accordingly.

\bibliography{references}
\bibliographystyle{IEEEtran}
\begin{IEEEbiography}[{\includegraphics[width=1in,height=1.25in,clip,keepaspectratio]{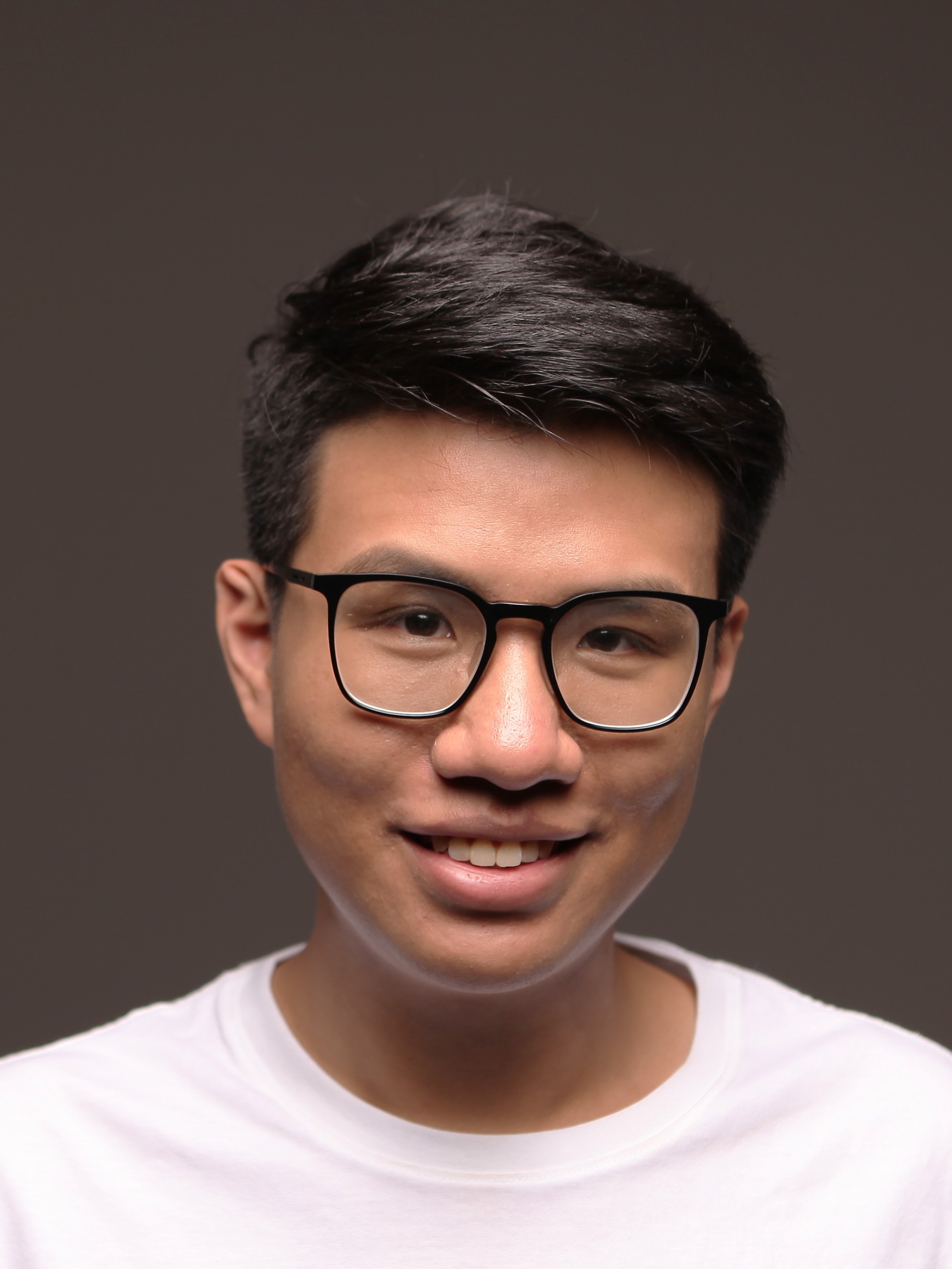}}]{Kai Tian}
 received the M.Sc. degree in automotive engineering from Chongqing University, China, 2019. He is now a PhD student at the Institute of Transport Studies, University of Leeds, UK, working from 2019. His main research interests include pedestrian-automated vehicle interaction, human factors and safety, and decision-making modelling.\end{IEEEbiography}

\begin{IEEEbiography}[{\includegraphics[width=1in,height=1.25in,clip,keepaspectratio]{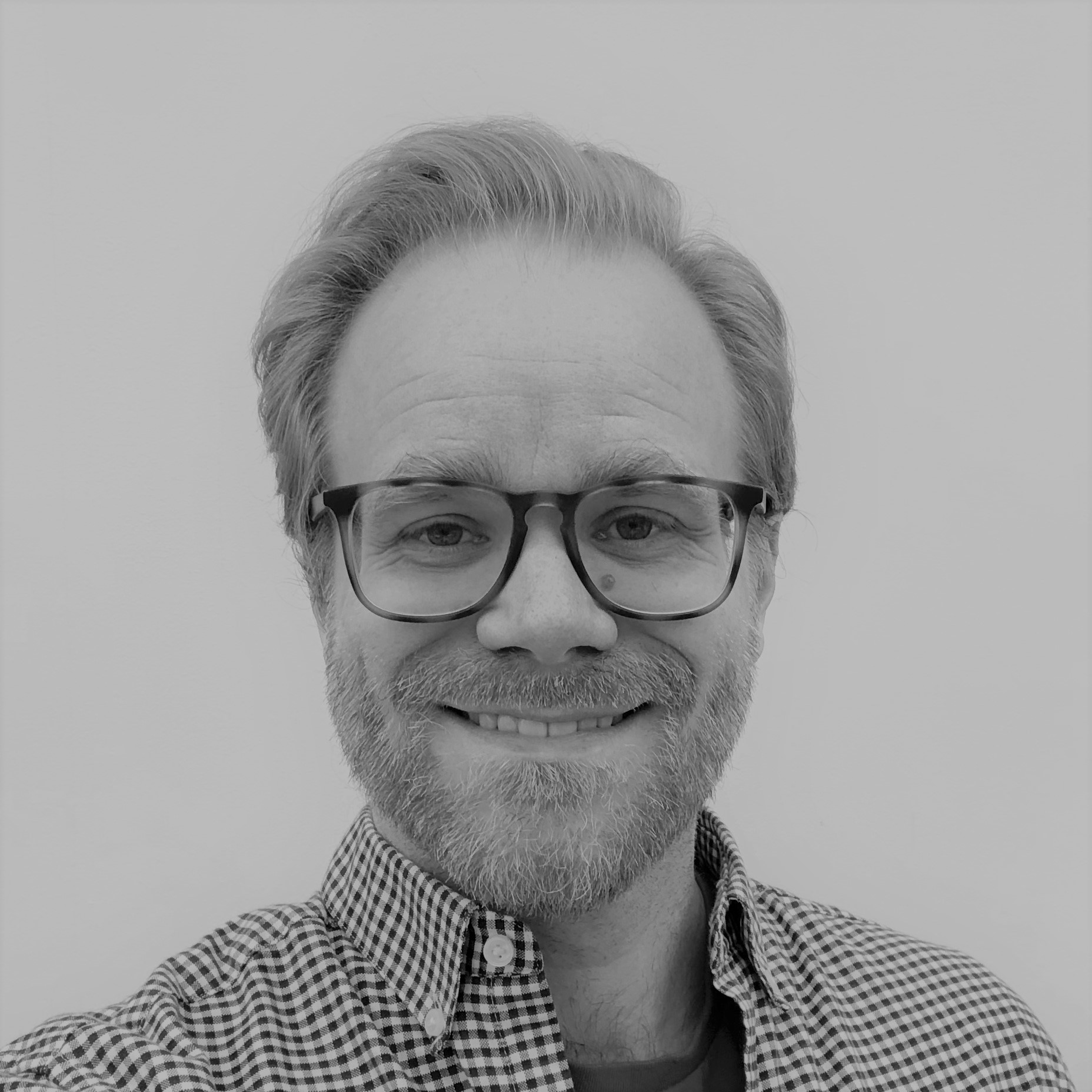}}]{Gustav Markkula}
received the M.Sc. degree in engineering physics and complex adaptive systems and the Ph.D. degree in machine and vehicle systems from Chalmers University of Technology, Gothenburg, Sweden, in 2004 and 2015, respectively. After having worked in the automotive industry for more than a decade, he is now Chair in Applied Behaviour Modelling at the Institute for Transport Studies, University of Leeds, U.K. His main research interests include quantitative, cognitive modeling of road user behavior and interaction, and virtual testing of vehicle technology and automation.\end{IEEEbiography}

\begin{IEEEbiography}[{\includegraphics[width=1in,height=1.25in,clip,keepaspectratio]{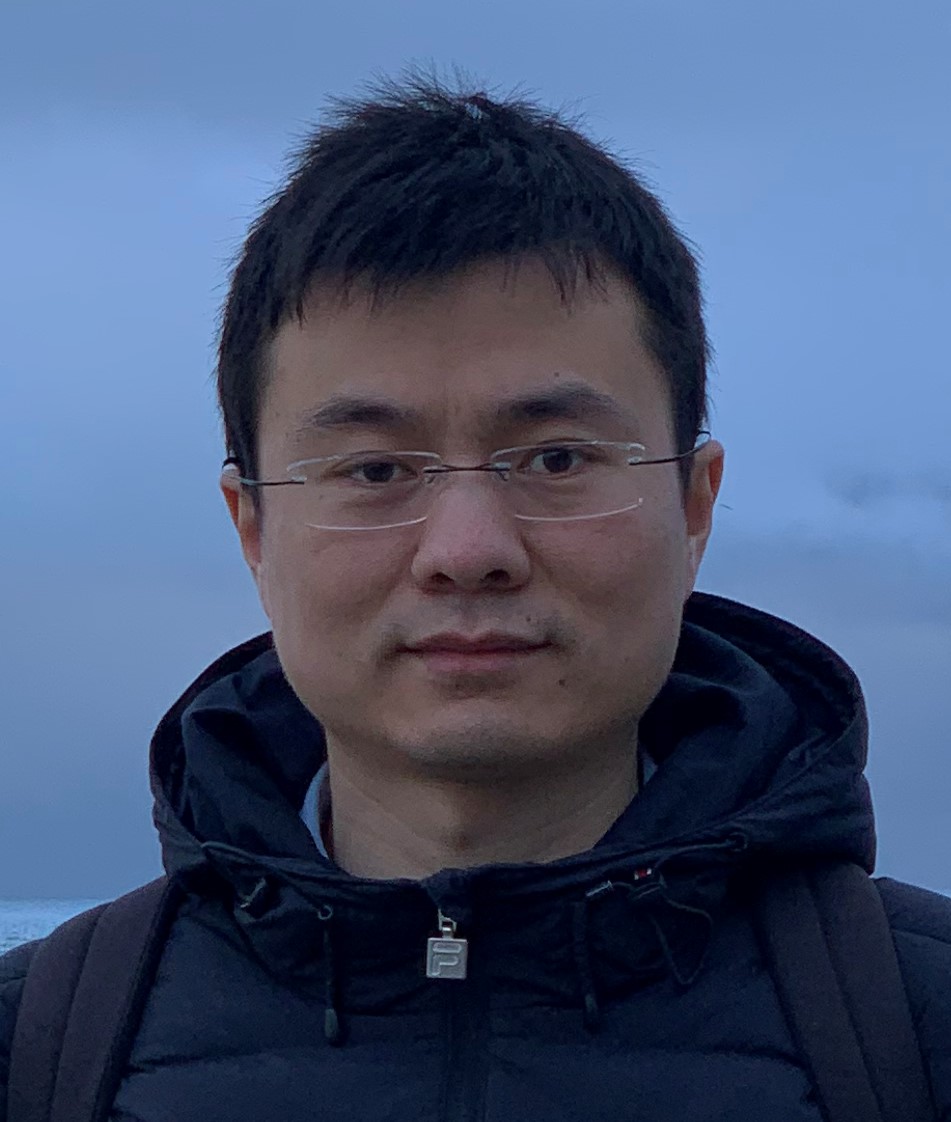}}]{Chongfeng Wei}
 received the PhD degree in mechanical engineering from the University of Birmingham, UK, in 2015. He is now an assistant professor at Queen's University Belfast, UK. His current research interests include decision making and control of intelligent vehicles, human-centric autonomous driving, cooperative automation, and dynamics and control of mechanical systems. He is also serving as an Associate Editor of IEEE Open Journal of Intelligent Transportation Systems, IEEE Transactions on Intelligent Vehicles, and IEEE Transactions on Intelligent Transportation Systems.\end{IEEEbiography}

\begin{IEEEbiography}[{\includegraphics[width=1in,height=1.25in,clip,keepaspectratio]{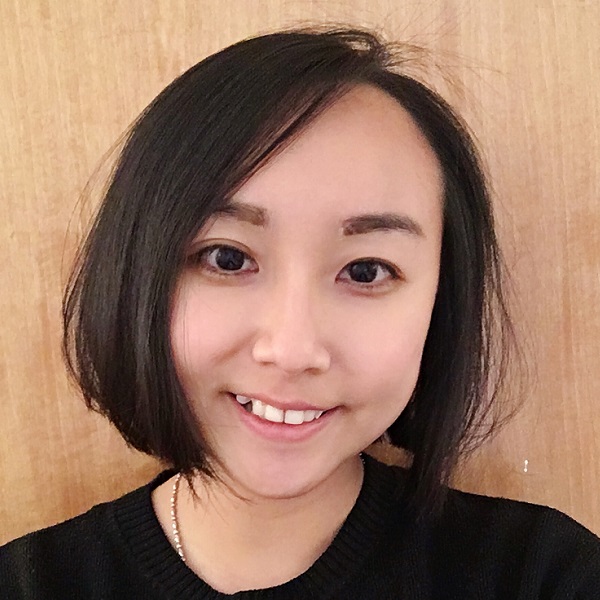}}]{YeeMun Lee}
 is currently a senior research fellow at the Institute for Transport Studies, University of Leeds. She obtained her BSc (Hons) in Psychology and her PhD degree in driving cognition from The University of Nottingham Malaysia in 2012 and 2016 respectively. Her current research interests include investigating the interaction between automated vehicles and other road users using various methods, especially virtual reality experimental designs. Yee Mun is involved in multiple EU-funded projects and is actively involved in the International Organisation for Standardisation (ISO).\end{IEEEbiography}

\begin{IEEEbiography}[{\includegraphics[width=1in,height=1.25in,clip,keepaspectratio]{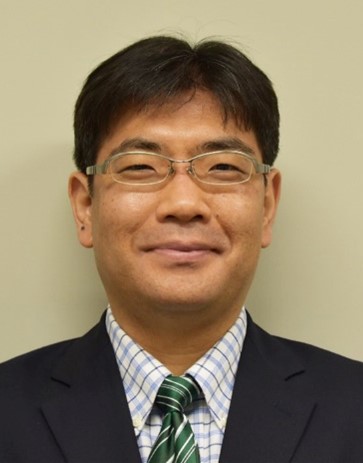}}]{Toshiya Hirose}
 received the master’s degrees and the Ph.D. degree from Shibaura Institute of Technology, Tokyo, Japan in 2002 and 2005. He is currently an Associate Professor with the Department of Engineering Science and Mechanics, Shibaura Institute of Technology. Before joining Shibaura Institute of Technology, he has worked with the National Traffic Safety and Environment Laboratory in Japan, and he was in charge of developing safety regulations for vehicles. He belonged to the Intelligent Transport Studies, the University of Leeds as the visiting researcher from 2019 to 2020. His active research interests include autonomous vehicles, driver assistance systems, active safety, driving simulators and human behavior models.\end{IEEEbiography}

\begin{IEEEbiography}[{\includegraphics[width=1in,height=1.25in,clip,keepaspectratio]{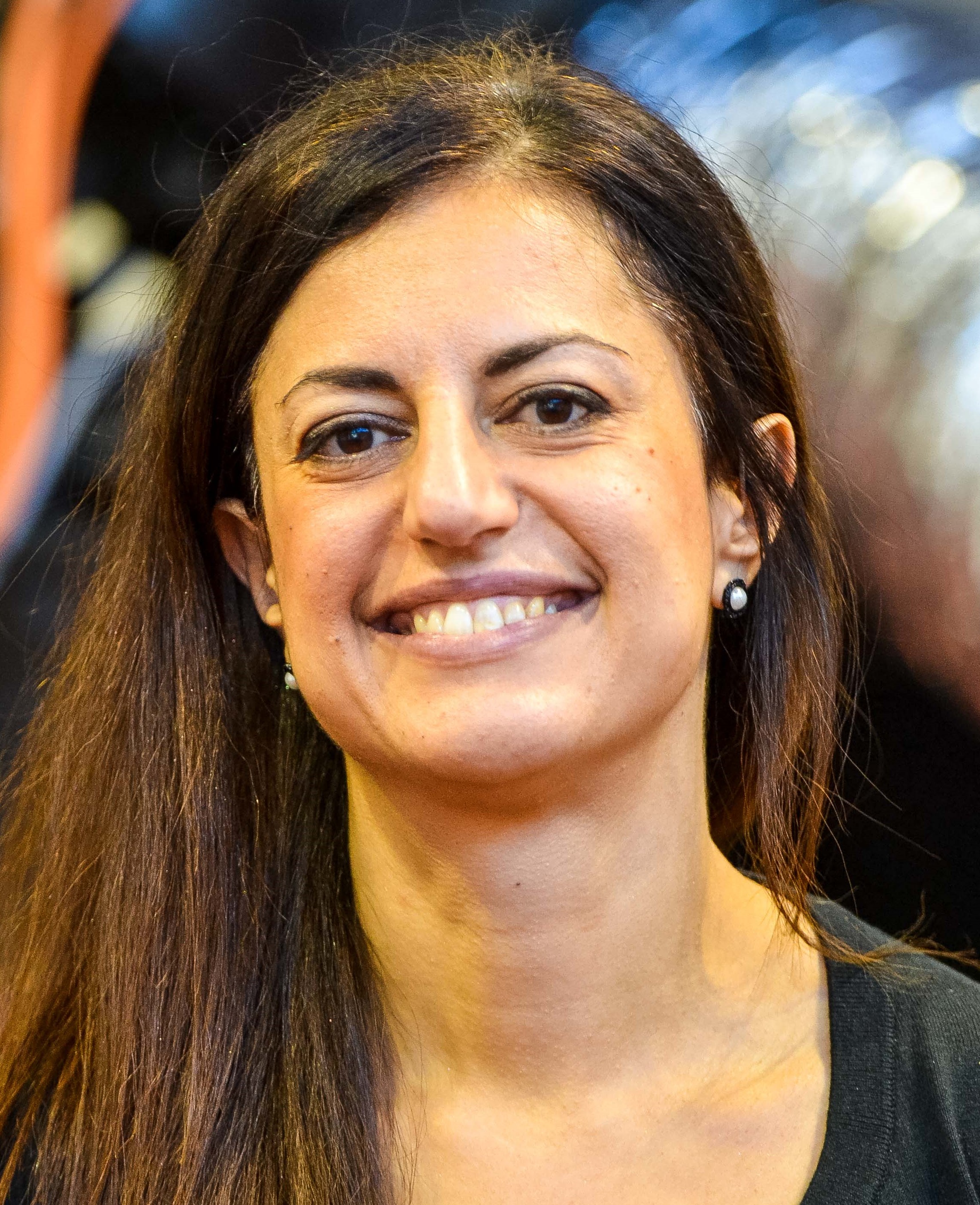}}]{Natasha Merat} is a Professor of Human Factors and Transport Systems at ITS, University of Leeds. She is leader of the multidisciplinary Human Factors and Safety Group and academic lead of Virtuocity at Leeds. She has a PhD in Psychology from Leeds, and her research interests are in understanding user interaction with new technologies in transport.
\end{IEEEbiography}

\begin{IEEEbiography}[{\includegraphics[width=1in,height=1.25in,clip,keepaspectratio]{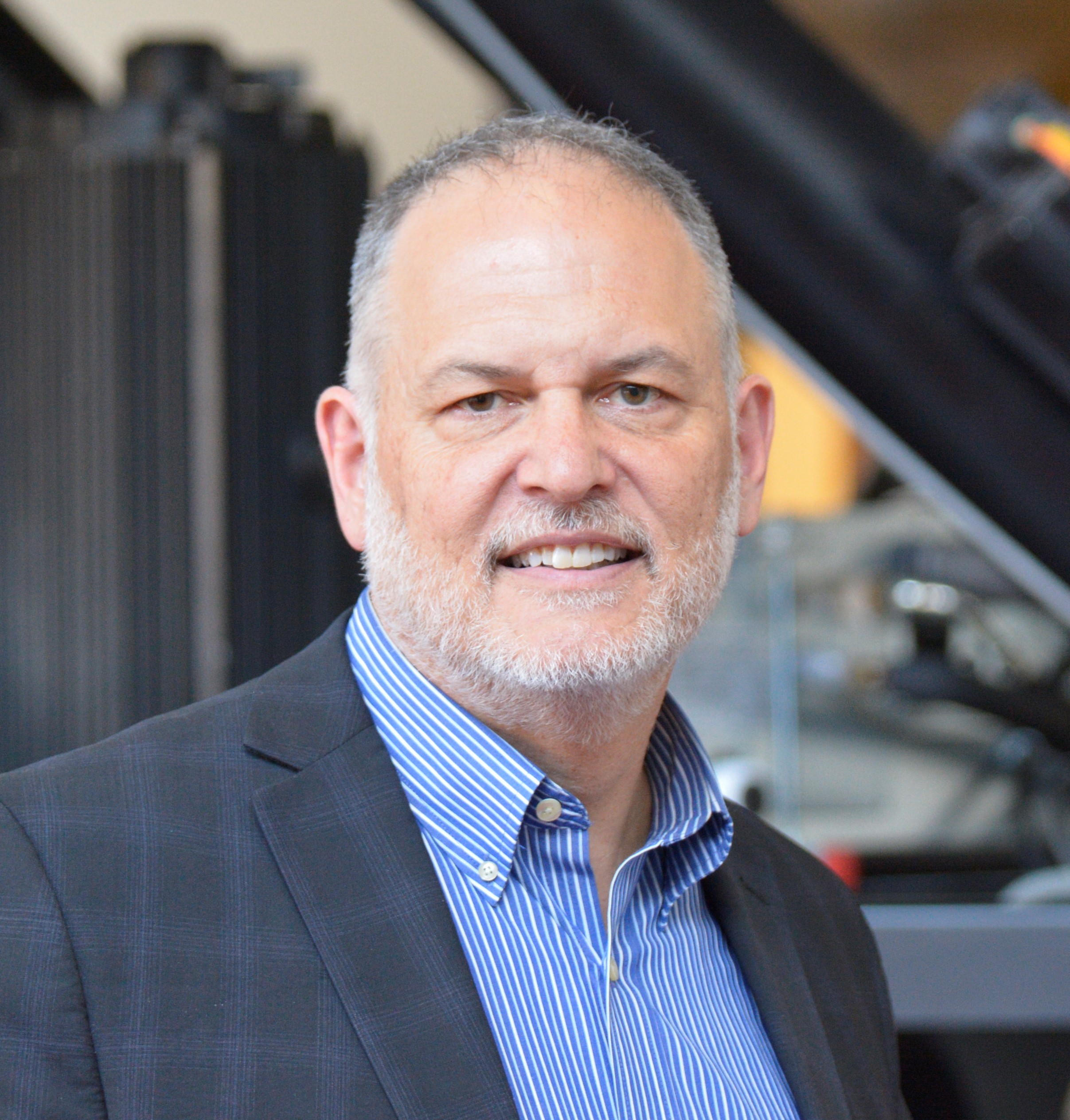}}]{Richard Romano} has over thirty years of experience developing and testing AVs and ADAS concepts and systems which began with the Automated Highway
Systems (AHS) project while he directed the Iowa Driving Simulator in the early 1990’s. He received his BASc and MASc in Engineering Science and Aerospace Engineering respectively from the University of Toronto, Canada and a PhD in Motion Drive Algorithms for Large Excursion Motion Bases, Industrial Engineering from the University of Iowa, USA. In addition to a distinguished career in industry he has supervised numerous research projects and
authored many journal papers. In 2015 he was appointed Leadership Chair in Driving Simulation at the Institute for Transport Studies, University of Leeds, UK. His research interests include the development, validation and application of transport simulation to support the human-centered design of vehicles and infrastructure.
 \end{IEEEbiography}

\clearpage
\renewcommand{\thefigure}{A.\arabic{figure}}
\setcounter{figure}{0}
\appendices

  \section{Supplementary file}
  \subsection{Simulation tool}
  \label{FirstAppendix}

\begin{figure}[htbp]
      \centering
     \includegraphics[scale=1]{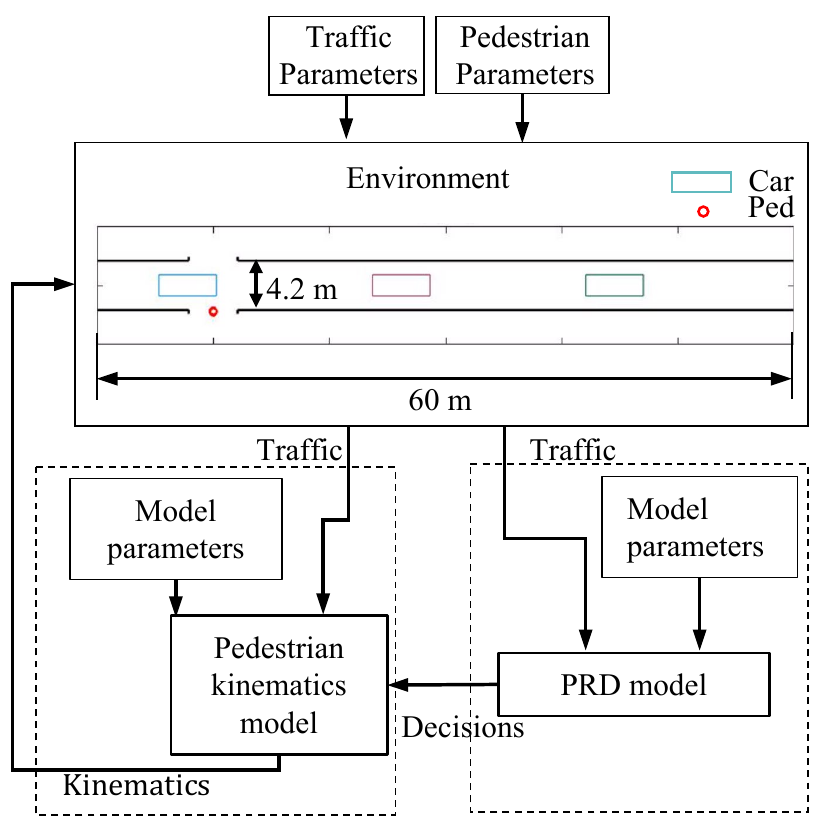}
      \caption{Structure of the simulation tool. The traffic environment contains a single lane (60 m long and 4.2 m wide) and a fleet of vehicles (colored rectangles).}
      \label{fiflowchart}
\end{figure}

In this study, an agent-based simulation tool is proposed using the established PRD models for reproducing pedestrian crossing behavior at uncontrolled intersections with traffic flow. The framework mainly includes three parts: PRD model, environment model, and pedestrian kinematics model (Fig.\ref{fiflowchart}).  The detailed process of the simulation tool is as follows:

\noindent (i) Generate the traffic environment using the given traffic and pedestrian parameters.

\noindent (ii) Generate a pedestrian agent at a random location on the pavement near the crosswalk. After that, the pedestrian walks to the edge of the pavement. Since this study focuses on the crossing decisions in the traffic flow, the pedestrian performs the PRD model after the first vehicle has passed him/her (Algorithm. \ref{alg1}).

\noindent (iii) The PRD model generates each pedestrian's decision and initiation time through a Monte Carlo sampling method (Algorithm. \ref{alg2}).

\noindent (iv) Pedestrians cross the road and walk to the opposite side of the road. The simulation model stops when the traffic scenario ends or all pedestrians cross the road.

A demonstration video of the simulation tool is also provided. Please see the attachment.

\begin{algorithm}
\caption{ Simulation model based on the model}
\hspace*{0.02in} {\bf Input:} 
Model parameters $\rho_{0},\rho_{1},\rho_{2},\rho_{3},\beta_{1},\beta_{2},\beta_{3},\beta_{4}, b$\\
\hspace*{0.02in} {\bf Output:}
$\boldsymbol{u}, \boldsymbol{t}_{int}$
\begin{algorithmic}[1]
\STATE $I_{r}=I$ \COMMENT{ Number of remaining participants $I_{r}$ and total number participants $I$ }
\FOR{$n$th gap in traffic $N$} 
    \STATE $\dot{\theta}_{n} \leftarrow \text{Eq}.\ref{1}$
    \STATE $X_{1,n}, X_{1,n} \leftarrow \text{Eq}.\ref{2}$
    and $\text{Eq}.\ref{3}$
    \STATE $p_{n} \leftarrow \text{Eq}.\ref{5}$ 
    \STATE $P_{n}=p_{n} \cdot (1-P_{n-1}) \leftarrow \text{Eq}.\ref{9}$
    \FOR{$i$th pedestrian \text{in} $I_{r}$}
        \STATE  $u_{i}$ $\leftarrow$ $Binomial(1,P_{n,i})$ \COMMENT{Sampling: crossing decision} 
        \IF {$u_{i}==1$}
            \STATE $f(t_{int})$ $\leftarrow$ $\text{Eq}.\ref{9}$ $\text{or}$ $\text{Eq}.\ref{10}$ \COMMENT{Caulculate probability density function of crossing decision} 
            \STATE $t_{int,i}$ $\leftarrow$ Algorithm. \ref{alg2}                     \COMMENT{Sampling: crossing initiation time} 
        \ELSE
            \STATE Continue
        \ENDIF
    \ENDFOR
    \STATE $I_{r}=I_{r}-\text{length}(\boldsymbol{t}_{int})$ \COMMENT{Update remaining participants}
\ENDFOR
\end{algorithmic}
\label{alg1}
\end{algorithm}

\begin{algorithm}
\caption{Monte Carlo sampling of the model}
\hspace*{0.02in} {\bf Input:} 
$f(t_{int})$ \\
\hspace*{0.02in} {\bf Output:}
$t_{int,i}$
\begin{algorithmic}[1]
\STATE Initialise $s=1$
\WHILE{$s\neq 2$}
\STATE $\pi(x)=f(x)$
\STATE $s \leftarrow  \text{Uniform}(0,1) $;  
\STATE $y\leftarrow Q(x|y)$ \COMMENT{Arbitrary probability density}
\IF{$u\leq \text{min}(\frac{pi(y)Q(x|y)}{pi(x)Q(y|x)},1)$}
\STATE $t_{int,i}=y$
\STATE $s=s+1$
\ELSE
\STATE $s=1$
\ENDIF
\ENDWHILE
\end{algorithmic}
\label{alg2}
\end{algorithm}

\subsection{Detailed modeling results}
\label{SecondAppendix}

\begin{figure*}[htbp]
      \centering
     \includegraphics[scale=1.2]{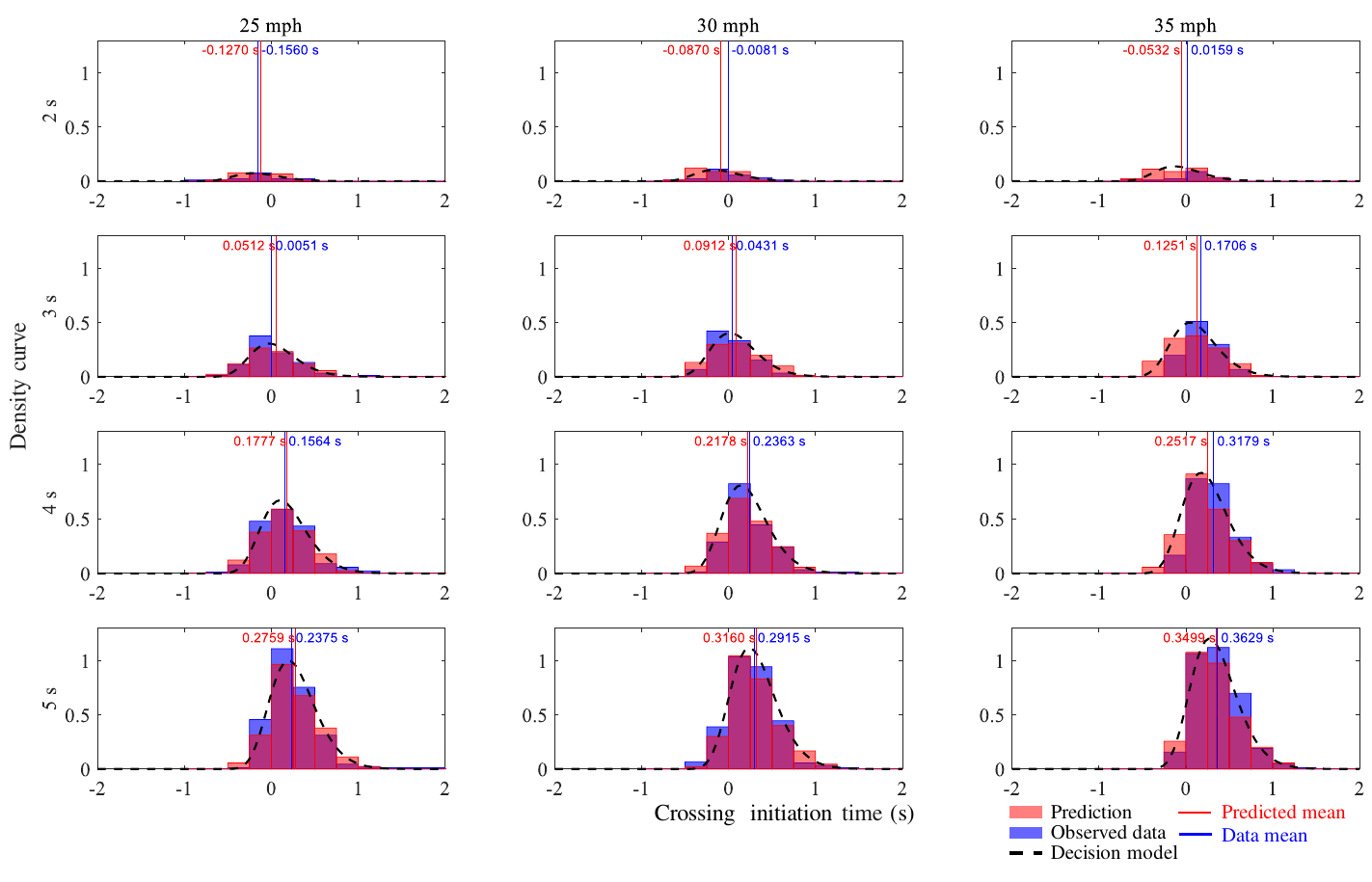}
      \caption{Predicted density function of crossing initiation time of the SW-PRD model based on dataset one. The predicted results, including density function, samplings and mean values of crossing initiation time, are compared with the observed data in terms of vehicle speed and traffic gap size.}
      \label{figlabela2}
\end{figure*}

\begin{figure*}[htbp]
      \centering
     \includegraphics[scale=1.3]{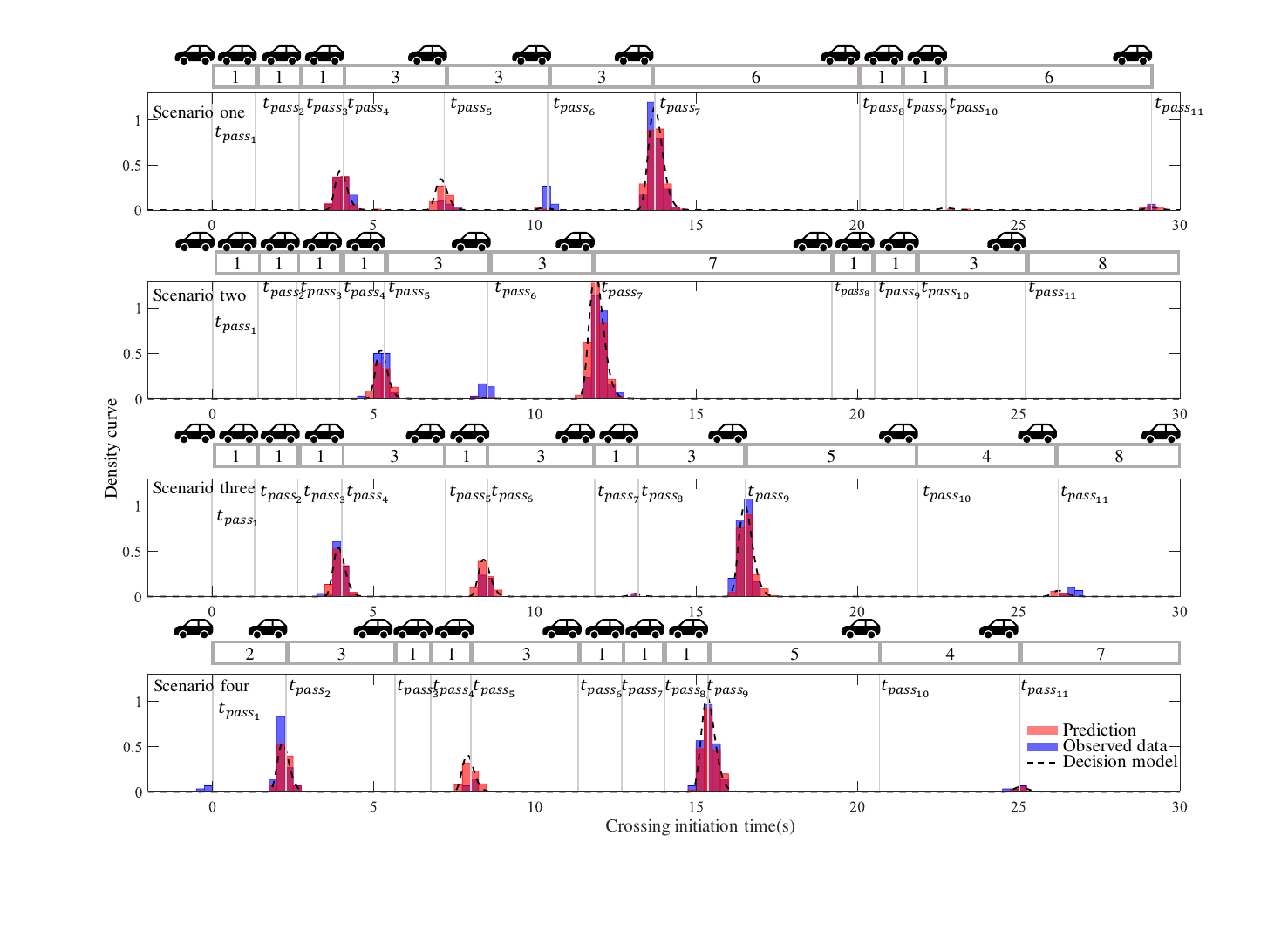}
      \caption{Predicted density function of crossing initiation time of the SW-PRD model based on dataset two. The predicted density functions and samplings are compared with the observed data. Regarding each traffic scenario, the order of traffic gaps is indicated above each sub-figure. The vertical lines represent the time when the rear end of the related vehicle passes the pedestrian's position, i.e., $t_{pass}$. }
      \label{figlabela3}
\end{figure*}

Detailed comparisons between modeling results and observations are shown in Fig.\ref{figlabela2} and Fig.\ref{figlabela3}. In Fig.\ref{figlabela2}, the probability density functions of crossing initiation time are plotted against time gaps and vehicle speeds. While, In Fig.\ref{figlabela3}, the probability density functions of crossing initiation time are plotted as a function of traffic scenarios and crossing initiation time.

\end{document}